\documentclass[]{aa}
\usepackage{graphicx}
\usepackage{txfonts}
\usepackage[round]{natbib}

\newcommand{\kms}{km\,s$^{-1}$}
\newcommand{\msun}{$M_\odot$}
\newcommand{\vsi}{$v \sin i$}
\newcommand{\te}{$T_{\rm eff}$}
\newcommand{\lgg}{$\log\,{g}$}
\newcommand{\bz}{\ensuremath{\langle \mathit{B}_z \rangle}}
\newcommand{\bs}{\ensuremath{\langle \mathit{B} \rangle}}
\newcommand{\ltsimeq}{\raisebox{-0.6ex}{$\,\stackrel
       {\raisebox{-.2ex}{$\textstyle <$}}{\sim}\,$}}

\begin{document}

\title{A Comprehensive Analysis of the Magnetic Standard Star HD
  94660: Host of a Massive Compact Companion?\thanks{Based in part on our own observations made with the
    European Southern Observatory (ESO) telescopes under the ESO
    programme 093.D-0367(A) and programmes 076.D-0169(A), 088.D-0066(A),
    087.D-0771(A), 084.D-0338(A), 083.D-1000(A) and 60.A-9036(A), obtained
    from the ESO/ST-ECF Science Archive Facility. It is also based in
    part on observations carried out at the Canada-France-Hawaii
    Telescope (CFHT) which is operated by the National Research
    Council of Canada, the Institut National des Science de
    l'Univers of the Centre National de la Recherche Scientifique of
    France and the University of Hawaii.}}

\author{J. D. Bailey\inst{1}
  \and
  J. Grunhut\inst{2}
  \and
  J. D. Landstreet\inst{3,4}
}

\institute{Max Planck Insitut f\"{u}r Extraterrestrische Physik, Giessenbachstrasse 1, 85748 Garching, Germany\\
\email{jeffbailey@mpe.mpg.de}
\and
ESO, Karl-Schwarzschild-Strasse 2, 85748 Garching, Germany
\and
Armagh Observatory, College Hill, Armagh, BT61 9DG, Northern Ireland, UK
\and
Department of Physics and Astronomy, The University of Western Ontario, London, Ontario, N6A 3K7, Canada
}

\date{Received 2014; accepted 2014}

\abstract{}{Detailed information about the magnetic geometry, atmospheric abundances and radial velocity variations has been obtained for the magnetic standard star HD~94660 based on high-dispersion spectroscopic and spectropolarimetric observations from the UVES, HARPSpol and ESPaDOnS instruments. }{We perform a detailed chemical abundance analysis using the spectrum synthesis code \textsc{zeeman} for a total of 17 elements. Using both line-of-sight and surface magnetic field measurements, we derive a simple magnetic field model that consists of dipole, quadrupole and octupole components. }{The observed magnetic field variations of HD~94660 are complex and suggest an inhomogeneous distribution of chemical elements over the stellar surface. This inhomogeneity is not reflected in the abundance analysis, from which all available spectra are modelled, but only a mean abundance is reported for each element. The derived abundances are mostly non-solar, with striking overabundances of Fe-peak and rare-earth elements. Of note are the clear signatures of vertical chemical stratification throughout the stellar atmosphere, most notably for the Fe-peak elements. We also report on the detection of radial velocity variations with a total range of 35~\kms\ in the spectra of HD~94660. A preliminary analysis shows the most likely period of these variations to be of order 840~d and, based on the derived orbital parameters of this star, suggests the first detection of a massive compact companion for a main sequence magnetic star.}{HD~94660 exhibits interestingly complex magnetic field variations and remarkable radial velocity variations. Long term monitoring is necessary to provide further constraints on the nature of these radial velocity variations. Detection of a companion will help establish the role of binarity in the origin of magnetism in stars with radiative envelopes. }

\titlerunning{The Magnetic Standard HD~94660}
\maketitle

\section{Introduction}
Of order 10\% of main sequence A and B stars are magnetic. For the majority of these stars, their magnetic properties are studied using circular polarisation spectra, which provide a measure of the mean magnetic field projected along the line-of-sight (\bz). A fraction of these stars rotate slowly and/or have strong enough magnetic fields that Zeeman splitting is observed in individual lines of the intensity (Stokes I) spectra \citep[see][and references therein]{Bailey2014}. The magnitude of this splitting provides a measure of the mean magnetic field modulus at the stellar surface (\bs). These magnetic A and B stars exhibit anomalous atmospheric abundances and are referred to as the chemically peculiar A-- type (Ap) stars. These stars often have \bz\ in excess of about 1~kG and \bs\ of order several thousands to tens of thousands of gauss \citep[e.g.][]{DL2009}.  

The chemical peculiarities of Ap stars can be quite striking. For example, the Fe-peak element Cr may be as much as 10$^2$ times overabundant compared to the Sun. Still more impressive are the abundances of rare-earth elements, which are commonly in excess of solar values by as much as 10$^4$ times. 

Ap stars may be highly variable, with their magnetic field strengths and spectral line strengths and shapes varying with the rotation period of the star. This variability is best explained via the rigid rotator model \citep[see][]{Stibbs1950}. In this model, the line-of-sight and magnetic axis are at angles $i$ and $\beta$ to the rotation axis, respectively. Therefore, different magnetic field measurements throughout the rotation cycle of the star are the result of observing the field at different orientations and since chemical elements are distributed non-uniformly and non-axisymmetrically over the stellar surface, spectrum variability is also observed \citep{Ryab1991}.

HD~94660 ( = HR~4263) is a bright ($V =$ 6.11) chemically peculiar
magnetic Ap star that is commonly used as a magnetic standard to test polarimetric systems in the southern hemisphere. The field was first discovered by \citet{BL1975} using an H$\alpha$ magnetograph; they reported a value for the line-of-sight magnetic field of \bz\ $= -3300 \pm 510$~G. HD~94660 is a sharp-lined star with clearly resolved Zeeman splitting in several spectra lines \citep{Mathys1990}. \citet{BLT1993} report a projected rotational velocity of \vsi\ $<$ 6~\kms and a roughly constant line-of-sight magnetic field strength of \bz\ $= -2520$~G, based on measurements using H$\beta$. A period of rotation of order 2700~d was first proposed by \citet{Hensberge1993} and later discussed by \citet{Mathys1997} with respect to \bs\ data. More recently, \citet{LBF2014} studied the field variations of HD~94660 from 17 FORS1 observations taken over a 6 year period from 2002 to 2008. They find a peak-to-peak variation in \bz\ of about 800~G, ranging from about $-2700$ and $-1900$~G. From these variations, a rotation period of 2800 $\pm$ 250~d is deduced, which agrees with previous determinations.

\citet{LM2000} were the first to report clear \bz\ variations (between about $-1800$ and $-2100$~G) which, together with \bs\ data from \citet{Mathys+Hubrig1997}, enabled them to model the magnetic field of this star using a colinear multipole expansion.
They adopted a model that consists of dipole, quadrupole and octupole
components with surface polar field strengths of $-8400$, $2700$ and $6900$~G, respectively. As seen
above, the value of \bz\ is always negative, indicating that $i +
\beta \ltsimeq 90^{\circ}$ i.e. only the negative magnetic pole
is observed. This is confirmed by this model where the determined
values for $i$ and $\beta$ are 5$^{\circ}$ and 47$^{\circ}$. We point
out that, in general, this model does provide a rough first
approximation to the field variations.

In this paper, we discuss efforts to model the magnetic field and chemical abundances of many elements based on high-dispersion, polarimetric spectra. The following section discusses the observations. Sect.~3 outlines the derived physical parameters; Sect.~4 discusses the magnetic field measurements and model; Sect.~5 and 6 describe the modelling technique and abundance analysis, respectively; Sect.~7 reports detected radial velocity variations; and Sect.~8 summarises the results of the paper. 

\begin{table}
\centering
\caption{Summary of the stellar and magnetic properties of HD~94660.}
\begin{tabular}{lrl}
\hline
Spectral type & A0p EuSiCr   &  \citealp{Rensonetal1991}  \\
$T_{\rm eff}$ (K) & 11300 $\pm$ 400 & This paper\\
log $g$ (cgs) & 4.18 $\pm$ 0.20  & This paper   \\
R (R$_\odot$) & 2.53 $\pm$ 0.37 & This paper\\
$v\sin i$ (km\,s$^{-1}$) & $<$2 & This paper \\
P (d) & 2800 $\pm$ 250 & \citealp{LBF2014}\\
$\log (L_\star/L_\odot)$ & 2.02 $\pm$ 0.10  & This paper\\
$M$ ($M_{\odot}$) & 3.0 $\pm$ 0.20 & This paper\\
$\pi$ (milliarcsec) & 6.67 $\pm$ 0.80 & van Leeuwen 2007\\ 
\hline
$B_{\rm d}$ (G) & $-7500$ & This paper\\
$B_{\rm q}$ (G) & $-2000$ & This paper\\
$B_{\rm oct}$ (G) & $7500$ & This paper\\
$i$ ($\degr$) & 16  & This paper\\
$\beta$ ($\degr$) & 30 & This paper\\
\hline\hline
\label{params}
\end{tabular}
\end{table}
\section{Observations}
We have acquired one spectropolarimetric observation of HD~94660 and have retrieved an additional seven archival spectra that were utilised in this study, all of which were taken with HARPSpol. This is a high-resolution ($R \simeq 115~000$), cross-dispersed echelle spectropolarimeter that covers a spectral range between 3780 -- 6910~\AA\ and is mounted on the European Southern Observatory (ESO) 3.6-m telescope located at La Silla. The high signal-to-noise ratio (SNR) observations consist of 6 circularly polarised Stokes $V$ spectra (with the 01-04-2012 observation also including linear polarisation in both Stokes $Q$ and $U$), and two unpolarised Stokes $I$ spectra, all acquired between May 2009 and April 2014. Each polarimetric observation (except for the observation taken on 31-05-2009) was obtained by acquiring four successive individual spectra with the quarter-wave plate rotated in such a way to acquire the desired Stokes spectrum \citep[see e.g.][ for further details]{rusomarov13}. 

The HARPSpol spectra were reduced using a modified version of the {\sc reduce} package \citep{piskunov02, makaganiuk11}. Wavelength calibration was performed using the spectrum of a ThAr calibration lamp and then corrected to the heliocentric rest frame. Normalisation of the spectra was achieved by first dividing the spectra by the optimally-extracted spectrum of the flat-field to correct for the blaze shape and fringing. The resulting spectra were then corrected by the response function derived from observations of the Sun. The last step involved fitting a smooth, slowly-varying function to the envelope of the entire spectrum. The final output is a set of continuum-normalised Stokes $I$ spectra, the Stokes parameter of interest ($V, Q, U$) and a diagnostic null spectrum that is calculated in such a way that the polarisation cancels out, which often allows us to identify spurious signals that are present in the processed data. The spectrum that was acquired on 31-05-2009 only completed half of the full polarimetric sequence, which enables the cancellation of first order (linear) wavelength drift, but does not correct for second order (quadratic) drift and does not allow the computation of a diagnostic null spectrum. We also note that the retarder angles listed in the observations obtained on 05-01-2010 were inconsistent with the usually adopted values for obtaining Stokes $V$ measurements and are likely erroneous. We therefore proceeded to reduce the data assuming the usual retarder angles for the sequence of observations, and then verified that the resulting polarised and unpolarised spectra were in good agreement with the other observations. 

In addition to the HARPSpol observations, we identified another archival, circularly polarised, high-resolution ($R \simeq 65~000$) spectropolarimetric observation acquired with ESPaDOnS on January 9, 2006. ESPaDOnS, which is mounted at the 3.6~m Canada-France-Hawaii Telescope (CFHT), is also a bench-mounted, cross-dispersed echelle spectropolarimeter, which covers a broader spectral range compared to HARPSpol of 3690 -- 10 481~\AA. The polarimetric observation was obtained by taking a sequence of four sub-exposures with different positions of the Fresnel Rhomb to acquire a single Stokes $V$ spectrum, according to the procedure described by \citet{donati97}. The ESPaDOnS spectra were processed using the automated reduction package {\sc libre-esprit}, following the double-ratio procedure \citep{donati97}. 

Lastly, we also found three nights of archival UVES data consisting of a total of 11 spectra. UVES is a cross-dispersed spectrograph mounted on the ESO 8.2-m Very Large Telescope (VLT) located at Paranal. It has both a blue and red arm offering different spectral resolutions and wavelength coverage. The blue arm  offers $R$ up to about 80\,000 and a spectral range of 3100 -- 4900~\AA, whereas the red arm covers 4800 -- $10\,200$~\AA\ with  $R$ up to about 110\,000. Table~\ref{speclog} summarises our entire collection of spectra for HD~94660.

\begin{center}
\begin{table*}
\caption{Log of available spectra for HD~94660. Successive columns list the instrument, date and JD of observation, exposure time, estimated signal-to-noise ratio (SNR) per 1.8~\kms\ velocity bin at 5000~\AA, spectral resolution ($R$) and wavelength coverage. }
\centering
\begin{tabular}{lccrrrr}
\hline\hline
Instrument & Date & JD & $t_{\rm exp}$ (s) & $SNR$ &  $R$ & $\lambda$~(\AA) \\
 & (DD-MM-YYYY) & (2450000+) & & & & \\
\hline
UVES & 01-05-2001 & 2031.464 & 145 & 254 & 65~030 & 3043 -- 3916 \\ 
     &            & 2032.464 & 70  & 252 & 74~450 & 4726 -- 6808 \\
     &            & 2031.466 & 70  & 266 & 74~450 & 4726 -- 6808 \\
UVES & 01-08-2001 & 2038.441 & 100 & 379 & 65~030 & 3731 -- 4999 \\
     &            & 2038.443 & 100 & 383 & 65~030 & 3731 -- 4999 \\
     &            & 2038.441 & 100 & 101 & 74~450 & 6650 -- 10~426\\
     &            & 2038.443 & 100 & 101 & 74~450 & 6650 -- 10~426 \\
UVES & 03-12-2005 & 3707.846 & 200 & 233 & 71~050 & 3044 -- 3917 \\
     &            & 3707.841 & 200 & 316 & 71~050 & 3281 -- 4563 \\
     &            & 3707.846 & 200 & 675 & 107~200 & 4726 -- 6835 \\
     &            & 3707.841 & 200 & 283 & 107~200 & 5708 -- 9464 \\
ESPaDOnS & 09-01-2006 & 3745.167 & 1200 & 270 & 65~000 & 3690 -- 10~481\\
HARPSpol & 24-05-2009 & 4975.546 & 600 & 311 & 115~000 & 3780 -- 6910\\
HARPSpol & 25-05-2009 & 4976.536 & 1200 & 398 & 115~000 & 3780 -- 6910\\
HARPSpol & 31-05-2009 & 4982.605 & 1200 & 476 & 115~000 & 3780 -- 6910\\
HARPSpol & 05-01-2010 & 5201.833 & 808 & 397 & 115~000 & 3780 -- 6910\\
HARPSpol & 19-05-2011 & 5701.450 & 800 & 392 & 115~000 & 3780 -- 6910\\
HARPSpol & 20-05-2011 & 5702.449 & 800 & 427 & 115~000 & 3780 -- 6910\\
HARPSpol & 01-04-2012 & 6018.559 & 1000 & 670 & 115~000 & 3780 -- 6910\\
HARPSpol & 28-04-2014 & 6775.610 & 360 & 320 & 115~000 & 3780 -- 6910\\
\hline\hline
\label{speclog}
\end{tabular}
\end{table*}
\end{center}

\section{Physical Parameters}
\subsection{Effective Temperature and Gravity}
Geneva and Str\"{o}mgren $uvby\beta$ photometry are available for
HD~94660 and both were utilised to determine the effective temperature \te\ and gravity \lgg\ for the star. For the Geneva photometry, we used the {\sc fortran} code developed by \citet{geneva}. A modified version of the \citet{NSW} code, that corrects the effective temperature to the Ap temperature scale \citep[see][for a complete discussion]{paper2}, was used for the Str\"{o}mgren photometry.

From the Geneva and Str\"{o}mgren photometric systems, we found
\te\ =  11500~K, \lgg\ = 4.10 and \te\ = 11100~K, \lgg\ = 4.26
respectively. We adopt the mean of these two sets of values for our
analysis with \te\ = 11300 $\pm$ 400~K and \lgg\ = 4.18 $\pm$ 0.2,
with the uncertainties estimated from the scatter between the
measurements, and by taking into account the intrinsic uncertainties in
computing these parameters for Ap stars.  These parameters are used
for our abundance analysis (see Sect.~6).

\subsection{Luminosity, Stellar Radius and Mass}
\citet{vanLee2007} reports a Hipparcos parallax of 6.67 $\pm$ 0.80
milliarcseconds. From this value, a distance to HD~94660 of about
150~pc is deduced. With a well determined distance, an appropriate
bolometric correction for Ap stars can be used to determine the stellar
luminosity \citep[see][]{paper2}, which is found to be $\log
L/L_{\odot} = 1.97 \pm 0.25$ with uncertainties propagated in the
usual way.   

Based on the \te\ and luminosity determinations, it is straightforward
to compute a stellar radius of $R = 2.53 \pm 0.37$~R$_{\odot}$. By
further comparing the position of HD~94660 to theoretical evolutionary
tracks \citep{Girardi2000} in an HR diagram, we are able to estimate
the star's evolutionary mass. Using the adopted uncertainties in \te\
and $L$, we proceed by comparing the star's position to multiple
evolutionary tracks of varying masses. In this manner, we estimate
that HD~94660 has a mass of about 3.0 $\pm$ 0.20~M$_{\odot}$. This is
in agreement to the mass of 3.51 $\pm$ 0.64~M$_{\odot}$ that we can estimate from the stellar radius and photometrically determined \lgg. Further, the position in the HR diagram suggests that HD~94660 has completed less than half of its main sequence lifetime. This is consistent with \citet{paper3}, who find strong magnetic fields only in young Ap stars. 

\section{Magnetic Field}

\begin{center}
\begin{figure}
\centering
\includegraphics*[angle=-90,width=0.45\textwidth]{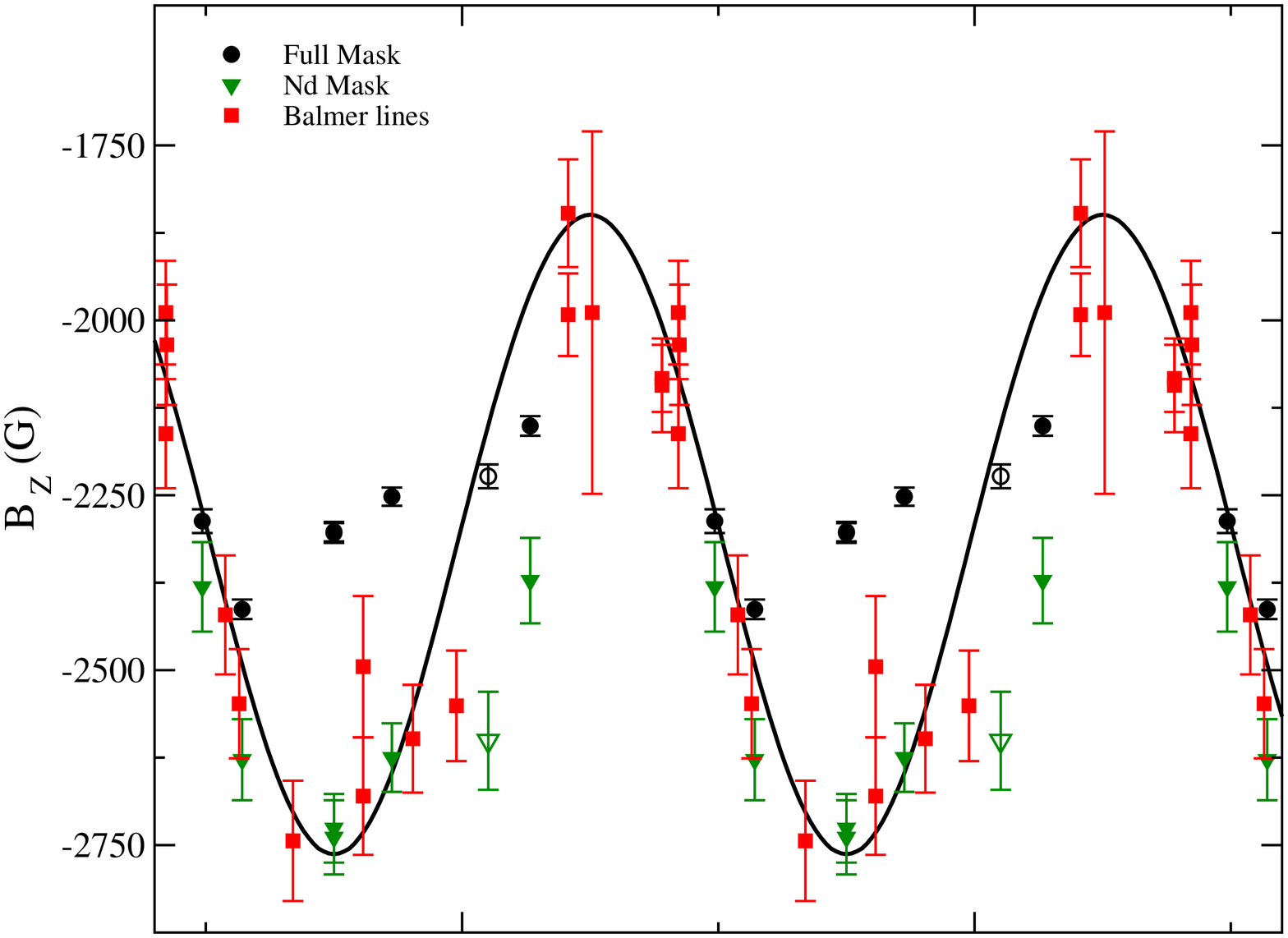}
\includegraphics*[angle=-90,width=0.45\textwidth]{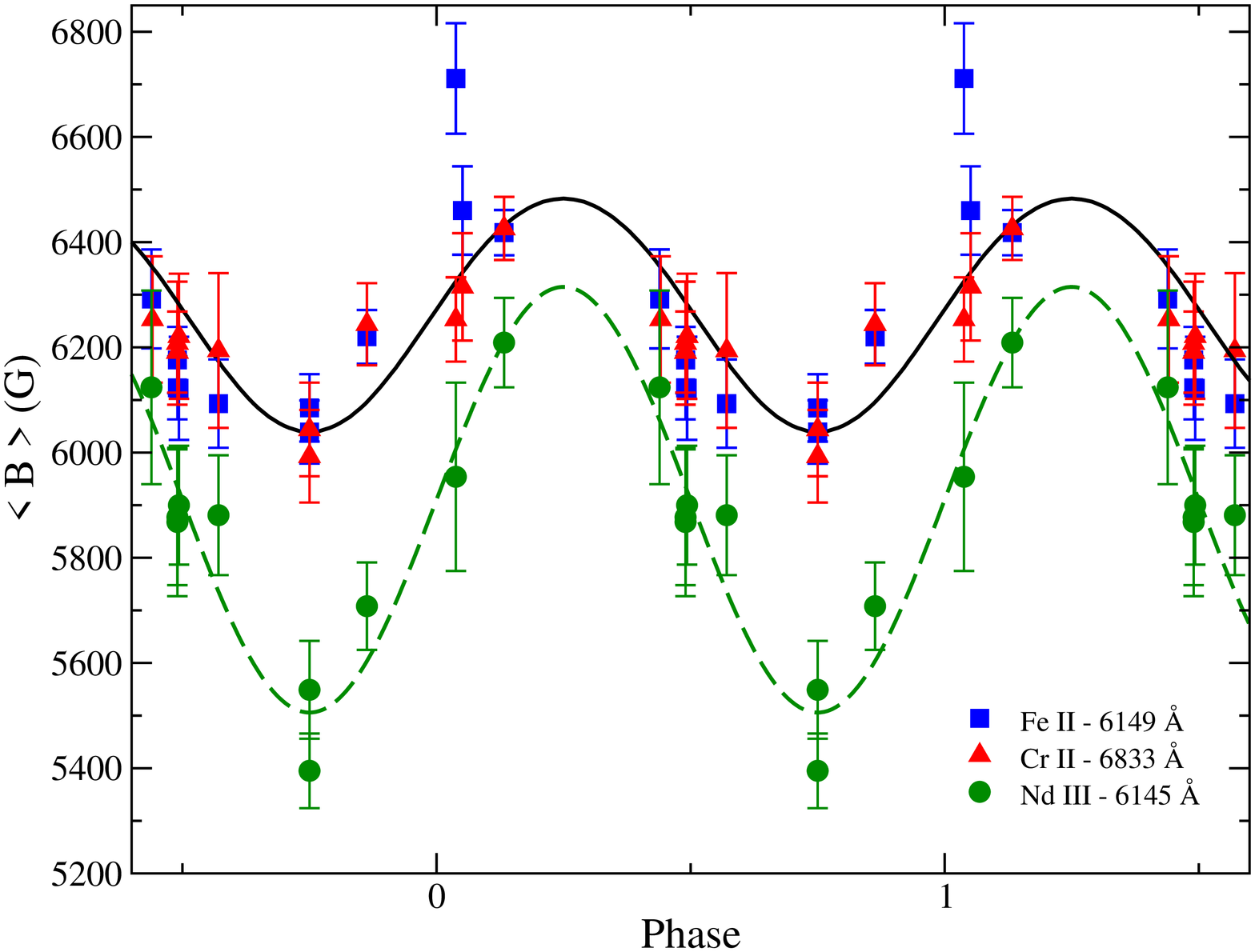}
\caption{\bz\ and \bs\ variations of HD~94660. The solid black lines are the adopted magnetic field model variations from Sect.~4.3. {\bf Top:} \bz\  LSD measurements using the full mask (black circles) and Nd mask (green triangles). Note that the open and filled symbols denote measurements from the ESPaDOnS spectrum and HARPSpol spectra, respectively.  Balmer line measurements from \citet{LBF2014} are the filled red squares. {\bf Bottom:} Phased \bs\ measurements from Fe~{\sc ii} 6149 (blue squares), Cr~{\sc ii} 6833 (red triangles) and Nd~{\sc iii} 6145 (green circles). The dotted green line is the best-fit sinusoidal variations to the Nd measurements.}
\label{magfield}
\end{figure}
\end{center}

\subsection{Longitudinal Magnetic Field Measurements}
For each of the Stokes $V$ spectra we measured the mean, surface-averaged longitudinal magnetic field (\bz) from line-averaged Least-Squares Deconvolved \citep[LSD; ][]{donati97} line profiles. The technique involves obtaining mean line profiles by combining all spectral lines in a given line list (normally metallic and He lines). The result is a much higher SNR with the ability to detect weaker Zeeman signatures due to magnetic fields. \bz\ was measured using the first-order moment method discussed by \citet{rees79}, and as implemented by \citet{donati97} and \citet{wade00} according to the equation:
\begin{equation}
\langle \mathit{B}_z \rangle = \frac{-2.14 \times 10^{11}}{\lambda z c }\frac{\int{(v - v_0)V(v){\rm d}v}}{\int{[1-I(v)]{\rm d}v}}.
\label{b_eq}
\end{equation}
In this equation, $v$ is the velocity within the LSD profile, $V(v)$ is the continuum-normalised Stokes $I$ profile and $I(v)$ is the continuum-normalised intensity profile. The wavelength $\lambda$ (in nm) and Land\'{e} factor $z$ correspond to the weighting factors used in the calculation of the LSD profiles (500-nm and 1.2, respectively). 

The LSD profiles were extracted using the iLSD code of \citet{kochukhov10}. As input the code requires a line mask that was extracted from the Vienna Atomic Line Database \citep[VALD;][]{vald1,vald2,vald3,vald4} for the spectral range covered by HARPS, using the mean abundances determined in this work and discussed in Sect.~\ref{abund_sect}. This linelist was used for both the HARPS and ESPaDOnS spectra to provide a consistent \bz\ measurement between the two datasets. We then proceeded to remove all lines that were blended with lines not used in this analysis, such as broad hydrogen lines or lines blended with strong telluric absorption bands. As a final step we then automatically adjust the line depths from their theoretical predictions to provide a best fit to the observed Stokes $I$ spectrum, while also removing lines that are poorly fit (such as those lines that show very strong Zeeman splitting). This final mask was then used to extract mean line profiles for all spectra that we label as `full mask'. We also extracted mean line profiles of individual elements for Si, Ti, Cr, Fe, and Nd, using the multi-profile capabilities of iLSD. In each case we provided two input masks based on our final full mask, one made entirely of the element of interest and the other containing all other lines. Both masks are used simultaneously as input and allow us to extract a representative mean, unblended profile of the element of interest. All LSD profiles were extracted onto a 1.8\,\kms\ velocity grid and uncertainties were computed by propagating the uncertainties in the final LSD profiles. 

Our final measurements are listed in
Table~\ref{bzmeas}. The top panel of Figure~\ref{magfield} plots the \bz\ measurements
against rotational phase for the full metallic and Nd masks as well as
the magnetic field model derived in Sect.~4.3. Also shown are the
Balmer line measurements of \citet{LBF2014}. Although the HARPSpol and ESPaDOnS magnetic data have not been intercalibrated for this star, we note that the consistency between the ESPaDOnS and HARPSpol measurements is satisfactory. Furthermore, previous studies have shown a good agreement between the polarimetric spectra acquired with ESPaDOnS and HARPSpol \citep{piskunov11}. 

\subsubsection{Measurements from individual spectral lines}
\begin{center}
\begin{figure}
\centering
\includegraphics*[angle=0,width=0.5\textwidth]{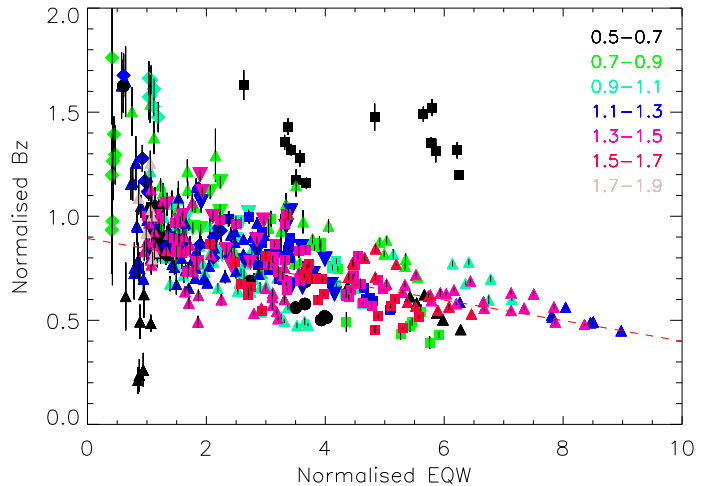}
\caption{The variations in magnetic field strength \bz\ versus EQW for individual spectral lines. Different symbols denote different elements, while different colours are used to represent different Land\'{e} factor ranges. Note that the absolute value of the field strength is shown and each measurement has been normalised as discussed in the text. The dashed line is a linear fit to the data.}
\label{eqw}
\end{figure}
\end{center}
The extremely sharp lines and strong magnetic field result in clean, strong Stokes $V$ signatures for individual spectral lines. This allows us to test the multi-line technique of LSD (see Sect.~4.1) by measuring \bz\ from individual lines of elements. We performed these measurements for a total of about 80 unblended lines which included the elements Si, Ti, Cr, Fe and Nd (the same elements for which we performed LSD to compute the field from individual elements). In general, the measurements agree with the values we derive and present in Table~\ref{bzmeas}, but with large scatter between different lines of an individual element. In an attempt to understand this scatter, we plot the measured \bz\ value against the equivalent width (EQW) for all the spectral lines (see Fig.~\ref{eqw}). To account for any variation in the field measurements due to the different phases, all values have been normalised by the \bz\ and EQW value for the Fe~{\sc i} 4982 line obtained for a given night. Note that the absolute value of the field strength is shown for clarity. 

The results show a clear trend, namely a stronger field strength 
for lines with smaller EQW and a weaker field strength for lines with 
larger EQW. We note that almost all of the outliers from this trend were 
lines with very small Land\'{e} factors. These results are difficult to 
interpret. On one hand, this result could reflect differential 
desaturation i.e. that strong lines should be more sensitive to the 
desaturation of the line due to magnetic splitting and so the relative 
change in the EQW of that line should be greater than for a weak line in 
the presence of a strong magnetic field. We attempted to test this 
hypothesis by producing a synthetic spectrum using {\sc zeeman} (see Sect.~5), which 
includes all of the measured spectral lines. Because the field strength 
is kept constant for a given phase in {\sc zeeman} for all lines, any 
systematic change in the measured \bz\ value as a function of the EQW 
can be attributed to differential saturation. Unfortunately, the results 
we derived from this test are also difficult to interpret, but it appears 
that the general trend of the synthetic tests do not agree with the 
results from the observations. Another possible explanation caused by 
differential saturation is that this result could simply reflect small 
systematic changes in the measured centre-of-gravity of Stokes $V$ (the 
numerator in Eqn.~\ref{b_eq}). This would predominantly affect strong 
lines with complex splitting patterns, resulting in a blended profile 
that does not have the same centre-of-gravity as the unsaturated line 
with the same splitting pattern. 

On the other hand, this result could be 
unrelated to saturation effects and could be a measure of the change of 
the decreasing field strength with increasing vertical height in the 
atmosphere: the cores of weak lines are formed deeper in the atmosphere 
and should have intrinsically stronger fields relative to strong lines 
with cores formed higher. Using our {\sc zeeman} model to 
estimate the depth of the atmosphere ($\sim$0.6\% of the stellar 
radius), we compute an expected change of the order of $\sim$4\% between 
the measured field strength of lines formed at the top and bottom of the 
atmosphere. Fig.~\ref{eqw} shows a typical variation in the strength of the magnetic field of order 30\% from weak to strong lines (closer to 50\% if you consider the extreme values). Therefore, it seems unlikely that this effect is important and the nature of this trend is still unclear. 

\begin{center}
\begin{table*}
\caption{Log of the line-of-sight surface field, \bz, measurements of
  HD~94660. Listed are the instrument, date and JD of observations, phase and \bz\
  measurements from the full stellar mask and only lines of Si, Ti,
  Cr, Fe and Nd.}
\centering
\begin{tabular}{lccccccccc}
\hline\hline
Instrument & Date & JD & Phase & \multicolumn{6}{c}{\bz\ (G)}\\
 & (DD-MM-YYYY) & (2450000+) & & Full Mask & Si & Ti & Cr & Fe & Nd\\
\hline
ESPaDOnS & 09-01-2006 & 3745.167 & 0.051 & $-2223 \pm 17$ & $-2343 \pm 77$ & $-1963 \pm 61$ & $-2174 \pm 25$ & $-2217 \pm 19$ & $-2601 \pm 70$\\
HARPSpol & 31-05-2009 & 4982.605 & 0.493 & $-2287 \pm 14$ & $-2231 \pm 71$ & $-2341 \pm 51$ & $-2114 \pm 22$ & $-2329 \pm 16$ & $-2381 \pm 64$\\
HARPSpol & 05-01-2010 & 5201.833 & 0.571 & $-2413 \pm 14$ & $-2460 \pm 70$ & $-2444 \pm 50$ & $-2273 \pm 23$ & $-2480 \pm 17$ & $-2628 \pm 58$\\ 
HARPSpol & 19-05-2011 & 5701.450 & 0.750 & $-2304 \pm 14$ & $-2515 \pm 68$ & $-2281 \pm 51$ & $-2106 \pm 22$ & $-2348 \pm 17$ & $-2739 \pm 53$\\
HARPSpol & 20-05-2011 & 5702.449 & 0.750 & $-2303 \pm 14$ & $-2502 \pm 68$ & $-2276 \pm 51$ & $-2092 \pm 22$ & $-2351 \pm 17$ & $-2726 \pm 52$\\
HARPSpol & 01-04-2012 & 6018.559 & 0.863 & $-2252 \pm 13$ & $-2399 \pm 65$ & $-2352 \pm 48$ & $-2151 \pm 22$ & $-2280 \pm 16$ & $-2625 \pm 49$\\
HARPSpol & 28-04-2014 & 6775.610 & 0.133 & $-2151 \pm 14$ & $-2305 \pm 71$ & $-1643 \pm 48$ & $-2149 \pm 21$ & $-2200 \pm 17$ & $-2372 \pm 61$\\
\hline\hline
\label{bzmeas}
\end{tabular}
\end{table*}
\end{center}

\subsection{Surface Magnetic Field Measurements}
\begin{center}
\begin{table*}
\caption{Log of surface field, \bs, measurements of HD~94660. For each spectra, the instrument, date and JD of observation, phase and \bs\ measurements from Nd~{\sc iii} $\lambda$6145, Fe~{\sc ii} $\lambda$6149 and Cr~{\sc ii} $\lambda$6833 are listed.}
\centering
\begin{tabular}{lccccrrr}
\hline\hline
Instrument & Date & JD & Phase & \multicolumn{3}{c}{\bs\ (G)} \\
&(DD-MM-YYYY) & (2450000+) & & Nd~{\sc iii} $\lambda$6145 & Fe~{\sc ii}
$\lambda$6149 & Cr~{\sc ii} $\lambda$6833 \\
\hline
UVES & 01-05-2001 & 2031.464 & 0.439 & 6124 $\pm$ 184 & 6292 $\pm$ 94 & --* \\ 
UVES & 01-08-2001 & 2038.441 & 0.441 & --* & --* & 6253 $\pm$ 120 \\
UVES & 03-12-2005 & 3707.841 & 0.038 & 5954 $\pm$ 179 & 6711 $\pm$ 105 & 6253 $\pm$ 80 \\
ESPaDOnS & 09-01-2006 & 3745.167 & 0.051 & -- & 6460 $\pm$ 84 & 6315 $\pm$ 102 \\
HARPSpol & 24-05-2009 & 4975.546 & 0.490 & 5877 $\pm$ 129 & 6123 $\pm$ 60 & 6208
$\pm$ 117 \\
HARPSpol & 25-05-2009 & 4976.536 & 0.490 & 5868 $\pm$ 141 & 6176 $\pm$ 63 & 6191 $\pm$ 77 \\
HARPSpol & 31-05-2009 & 4982.605 & 0.493 & 5900 $\pm$ 113 & 6122 $\pm$ 98 & 6221 $\pm$ 119 \\
HARPSpol & 05-01-2010 & 5201.833 & 0.571 & 5881 $\pm$ 114 & 6093 $\pm$ 84 & 6194 $\pm$ 147 \\
HARPSpol & 19-05-2011 & 5701.450 & 0.750 & 5395 $\pm$ 71 & 6085 $\pm$ 64 & 6044 $\pm$ 89\\
HARPSpol & 20-05-2011 & 5702.449 & 0.750 & 5549 $\pm$ 93 & 6039 $\pm$ 60 & 5993 $\pm$ 88\\
HARPSpol & 01-04-2012 & 6018.559 & 0.863 & 5708 $\pm$ 83 & 6220 $\pm$ 51 & 6244 $\pm$ 78 \\
HARPSpol & 28-04-2014 & 6775.610 & 0.133 & 6209 $\pm$ 85 & 6418 $\pm$ 43 & 6426 $\pm$ 60 \\
\hline\hline
\multicolumn{6}{p{0.5\textwidth}}{{\sc Notes.} (*) UVES spectra do not include these spectral lines (see Table~\ref{speclog})}\\
\label{bsmeas}
\end{tabular}
\end{table*}
\end{center}
The sharp-lined features and high resolving powers of the HARPSpol, UVES and
ESPaDOnS instruments allowed us to make measurements of the mean field
moduli (\bs) from observed Zeeman splitting in the lines of Nd~{\sc
  iii}~$\lambda$6145, Fe~{\sc ii}~$\lambda$6149 and Cr~{\sc
  ii}~$\lambda$6833. Notable exceptions are for the one ESPaDOnS
spectrum and two nights of UVES spectra.  The lower resolving power of ESPaDOnS compared to HARPSpol did
not allow us to see sufficient splitting in Nd~{\sc iii}, whereas insufficient wavelength coverage for the UVES spectra did not allow for measurements of splitting in Cr~{\sc ii} (on 01-05-2011) or Nd~{\sc iii} and Fe~{\sc ii} (on 01-08-2011) . 

To obtain values of \bs, the atomic data from the VALD database was used and the field strength calculated from the equation
\begin{equation}
\bs\ = \frac{\Delta\lambda}{4.67\times 10^{-13}\lambda_{o}^{2}z},
\label{Bs-eqn}
\end{equation}
where $\Delta\lambda$ denotes the shift of the $\sigma$ components
from the zero field wavelength, $\lambda_{o}$ is the rest wavelength in \AA\
and $z$ is the effective Land\'{e} factor of the line (1.00, 1.34 and
1.33 for $\lambda$6145, 6149 and 6833, respectively). The separation
of the elements of the Zeeman split lines that determine $\Delta\lambda$ was measured by fitting
Gaussians to each component using the {\it splot} function in
IRAF. The uncertainties were estimated by considering the dispersion
between a series of measurements of \bs\ from multiple Gaussian fits to
each component. \citet{Baileyetal2011} note that measurements of \bs\
from Eqn.~\ref{Bs-eqn} are very sensitive to small changes in
$\Delta\lambda$. Because of the high resolution and essentially zero rotation
rate of HD~94660, the dispersion in our measurements of $\Delta\lambda$
is less than about 0.005~\AA\ in most cases. Table~\ref{bsmeas}
summarises these \bs\ measurements and the bottom panel of
Fig.~\ref{magfield} show the phased rotational variations of the
surface fields from the three elements.

The scatter of \bs\ values around the mean curve confirms that our uncertainty estimates are reasonable. 
The consistency of measurements from each individual element taken on successive nights (24-05-2009
and 25-05-2009) and the agreement, within estimated uncertainties,
between \bs\ measurements of the Fe-peak elements Cr~{\sc ii} and
Fe~{\sc ii} (except for a single measurement) further suggests that the quoted uncertainties are realistic. The
surface field appears to vary sinusoidally; however, the
average surface field from the Fe-peak elements is of order 400~G
larger than that of Nd~{\sc iii}. Further, the peak-to-peak amplitude of the
variations is twice as large for Nd~{\sc iii} (of order 800~G) compared to the
Fe-peak elements (of order 400~G). The fact that different elements
sample the field in significantly different ways suggest that the abundance distributions of different elements, in addition to the field structure, are inhomogeneous
over the stellar surface. This is further supported by the discrepant
\bz\ measurements of different elements discussed above and listed in Table~\ref{bzmeas}.

\subsection{Magnetic Field Model}
Given the long rotation period of HD~94660, we are fortunate to have
sufficient phase coverage for both \bz\ and \bs\ in order to derive a
simple magnetic field model. In general, the FORS1 data show that the Balmer line and metallic line \bz\ measurements agree closely, except for one discrepant metal \bz\ value \citep{LBF2014}. However, a large discrepancy exists between all FORS1 measurements and those from high-resolution spectra. In particular, the FORS1 data show a much larger range in \bz. This can be seen in the top panel of Fig.~\ref{magfield} where the FORS1 Balmer line measurements vary by order 800~G whereas the full LSD mask from the high resolution spectra vary by about 300~G. We opt for using the Balmer line \bz\ variations as discussed by \citet{LBF2014} to derive a field geometry. We point out that this is the main reason for the discrepancy between our field structure (see below) and the one derived by \citet{LM2000}.

The {\sc fortran} program {\sc fldsrch.f} \citep{LM2000}
derives a magnetic field geometry based on the observed \bz\ and \bs\ variations. It produces a simple co-axial multipole expansion that consists of 
dipole ($B_{\rm d}$) , quadrupole ($B_{\rm q}$) and octupole ($B_{\rm oct}$) components with the angles between the
line-of-sight and rotation axis ($i$) and the magnetic field and
rotation axes ($\beta$), specified.  For HD~94660, a new
field geometry with $i = 16^{\degr}$, $\beta = 30^{\degr}$, $B_{\rm d}
= -7500$~G, $B_{\rm q} = -2000~G$, and $B_{\rm oct} = 7500$~G is
found. The phased \bz\ and \bs\ variations using our adopted geometry are
shown in Fig.~\ref{magfield} and are in good agreement with the
longitudinal field variations measured by \citet{LBF2014} from FORS1, as well as the measured Fe-peak surface field variations.   

\begin{center}
\begin{figure}
\centering
\includegraphics*[angle=-90,width=0.5\textwidth]{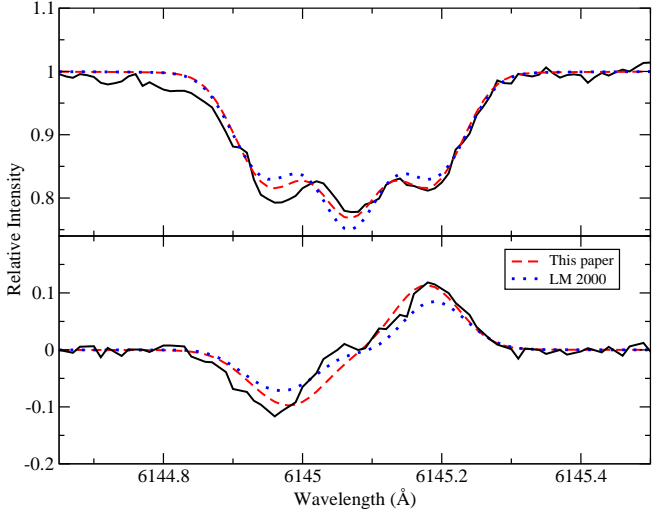}
\caption{A comparison of magnetic field models. In black are the
  observed Stokes $I$ (top) and $V$ (bottom) profiles for Nd~{\sc iii}
$\lambda$6145 for the 28-08-2014 HARPSpol spectrum. We overplot the computed line profiles for our adopted magnetic field geometry (red dashed line) and the magnetic field geometry of \citet{LM2000} (blue dotted line).}
\label{model}
\end{figure}
\end{center}

Although our magnetic field model is based on the fit to the large amplitude \bz\ FORS1 data, it is nevertheless useful to compare our geometry to the one from \citet{LM2000} because it is not entirely clear which, if either, is better to use. In general, our adopted field geometry is not drastically different than the one proposed by \citet{LM2000} (see Sect. 1). The angles $i + \beta$ are roughly the
same, although the individual values of these angles differ (recall
that $i$ and $\beta$ can be interchanged). The values of $B_{\rm d}$ and $B_{\rm oct}$ are similar; however, the relatively small $B_{\rm q}$, although comparable in magnitude, is negative in our derived geometry and positive in the geometry of \citet{LM2000}. The circular polarisation signatures in Stokes $V$ are also well produced with both
geometries. Fig.~\ref{model} compares the observed and synthetic Stokes
$I$ and $V$ profiles, using both models, for Nd~{\sc iii}
$\lambda$6145 (note that only one phase is shown but the splitting is equally well reproduced at all phases). It is obvious from this figure that both fields do an
adequate job at reproducing the observed magnetic properties in the
spectrum of HD~94660. Based on Figures~\ref{model}, it is unclear which model is best and both models are too coarse to describe the complex field variations over the visible stellar surface. Although our field model is preferred, we perform an abundance analysis using both geometries (see Sect.~5).

\begin{center}
\begin{figure}
\centering
\includegraphics*[angle=-90,width=0.5\textwidth]{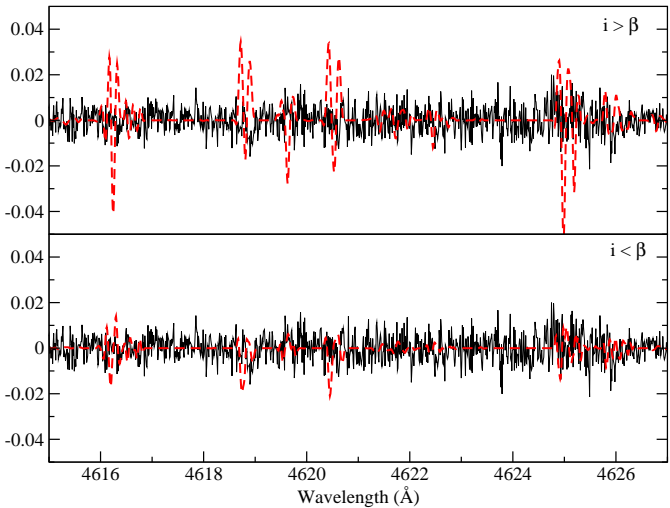}
\caption{A comparison of the Stokes Q signatures of synthetic models to the HARPSpol observation on 01-04-2012. In black is the
  observed Stokes $Q$ spectrum. Overplotted is our adopted magnetic field model (red dashed line). The top panel assumes $i > \beta$, whereas the bottom assumes $i < \beta$.}
\label{linpol}
\end{figure}
\end{center}

\citet{LM2000} suggest that $\beta$ is likely smaller than $i$ for a very large fraction of slow rotators, arguing that for long period stars ($P \ge 25$~d) it is equally likely for the rotation pole to be positioned anywhere on the visible hemisphere. For this reason, smaller values of $i$ are statistically unlikely, since it would require $i$ to be within a small region of the visible hemisphere. However, their geometries are based only on intensity and circular polarisation spectra which makes it difficult to distinguish $i$ and $\beta$. Fortunately, we have available two observations of the linear polarisation (Stokes $Q$ and $U$) spectra of HD~94660 taken with the HARPSpol instrument on 01-04-2012 (see Sect.~2). The linear polarisation data will allow us to distinguish which of the two angles is larger. Figure~\ref{linpol} compares the Stokes $Q$ signature for our adopted magnetic field model, assuming that $i > \beta$ (top panel) and $i < \beta$ (bottom panel). For $i > \beta$, the model clearly predicts a Stokes $Q$ signature that is too large compared to the observations. Alternatively, for $i < \beta$, the model predicts a much more modest signature, in relative agreement to what is observed, although we admit that the model does not accurately represent the true variations in linear polarisation for this star (we note that similar results are also seen with Stokes $U$ and with the \citet{LM2000} model). This figure strongly suggests that, contrary to the assumption of \citet{LM2000}, for HD~94660 $\beta$ is the larger of these two angles for both models. 
  
\section{Spectrum Synthesis Program}

\subsection{{\sc zeeman.f}}
The {\sc fortran} program {\sc zeeman.f} is a spectrum synthesis
program developed by \citet{Landstreet1988}. {\sc
  zeeman.f} performs line formation and radiative transfer in a
magnetic field. The input magnetic field geometry consists of the axisymmetric superposition of a dipole, quadrupole and octupole with specified values of the angles
$i$ and $\beta$.  {\sc zeeman.f} interpolates an appropriate atmospheric model from
a grid of solar abundance ATLAS9 models based upon the \te\ and \lgg\
supplied. The atomic data needed for line synthesis are taken from the VALD database. {\sc
  zeeman.f} assumes a homogeneous abundance distribution vertically throughout the
stellar atmosphere and computes all four Stokes parameters. {\sc zeeman.f} allows up to six abundance rings that are symmetric about the magnetic axis to be specified on the stellar surface, with each ring having equal extent in colatitude. Within each ring, a uniform abundance is assumed. For each modelled spectrum, a best-fit \vsi\ and radial velocity $v_{\rm R}$ are calculated.   

\subsection{Choice of magnetic field model}
One of the major difficulties in performing the abundance analysis of
a magnetic star is the inclusion of the effects of the magnetic
field on spectral line formation. This is, in part, due to the limited
number of tools that include this physics, but also because of the
inherent ambiguity in selecting an appropriate magnetic field
model. Section~4 highlighted this discrepancy with two apparently
adequate field models to explain the observed Stokes $I$ and $V$
profiles of HD~94660. Therefore, we compared the abundances derived from these two magnetic field geometries. We also tested the abundances with {\bf a} simple dipolar model in which the value of $B_{\rm d}$ was chosen to be approximately three times the largest observed \bz\ value (this leads to roughly the same extrema in the observed \bz\ and \bs\ values).

We note that the agreement between the different models is impressive. For any given set of fundamental parameters (i.e. \te\ and \lgg), the difference in the derived abundances between the three models are generally less than about 0.1~dex and no worse than about 0.2~dex. It is encouraging that the choice of model does not
drastically influence the final abundances. In fact, a very coarse model, consisting only of a dipole, can give accurate abundances, which suggests that the final abundances may not be very sensitive to the difference between these simple models and the real field distribution. Apparently, the inclusion
a magnetic field model that roughly explains the \bz\ and \bs\
variations is sufficient to provide an accurate analysis. This lends support to studies such as the one by \citet{BLB2014} for which simple dipolar models are used to quantify the evolution of atmospheric abundances in Ap stars. Since the abundances depend little on the magnetic field model, the following section reports only on abundances found using the field geometry we present in Sect.~4.3.

\section{Elemental Abundances}\label{abund_sect}
\begin{table}
\caption{The abundances are measured with respect to H, and are tabulated with their associated uncertainties. For reference, the solar abundance ratio and the number of spectral lines modelled are also shown. }
\centering
\begin{tabular}{lr|rr}
\hline\hline
Element & $\log (N_{\rm X}/N_{\rm H})$ & Solar & \# Lines \\
\hline
He & $<-2.20$ & $-1.07$ & 2\\
O & $-3.77 \pm 0.20$ & $-3.31$ & 3\\
Mg & $-5.03 \pm 0.15$ & $-4.40$ & 1\\
Si & $-3.21 \pm 0.30$ & $-4.49$ & 5\\
Ca & $-6.19 \pm 0.33$ & $-5.66$ & 1\\
Ti & $-5.55 \pm 0.50$ & $-7.05$ & 7\\
Cr & $-3.86 \pm 0.30$ & $-6.36$ & 30\\
Mn & $-5.67 \pm 0.20$ & $-6.57$ & 2\\
Fe & $-3.10 \pm 0.50$ & $-4.50$ & 28\\
Co & $-4.51 \pm 0.25$ & $-7.01$ & 8 \\
Ni & $-5.24 \pm 0.20$ & $-5.78$ & 1\\
Sr & $-7.78 \pm 0.40$ & $-9.13$ & 2\\
La & $-7.68 \pm 0.16$ & $-10.9$ & 2 \\
Ce & $-6.86 \pm 0.20$& $-10.42$ & 3\\
Pr & $-6.80 \pm 0.25$ & $-11.28$ & 4\\
Nd & $-6.54 \pm 0.30$ & $-10.58$ & 9\\
Eu & $-7.49 \pm 0.20$ & $-11.48$ & 3 \\
\hline\hline
\end{tabular}
\label{abund}
\end{table}

\begin{center}
\begin{figure}
\centering
\includegraphics*[angle=-90,width=0.5\textwidth]{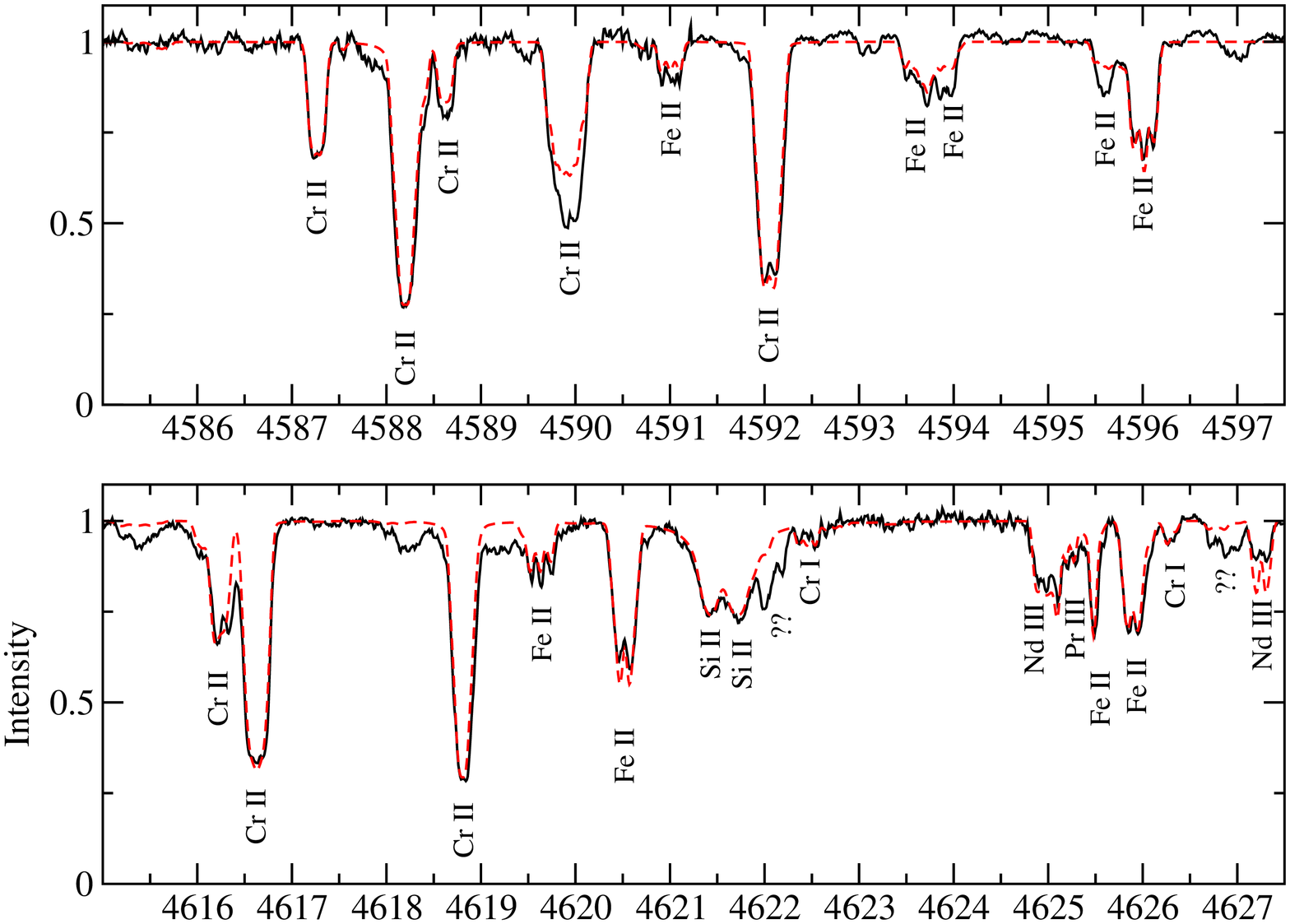}\\
\includegraphics*[angle=-90,width=0.5\textwidth]{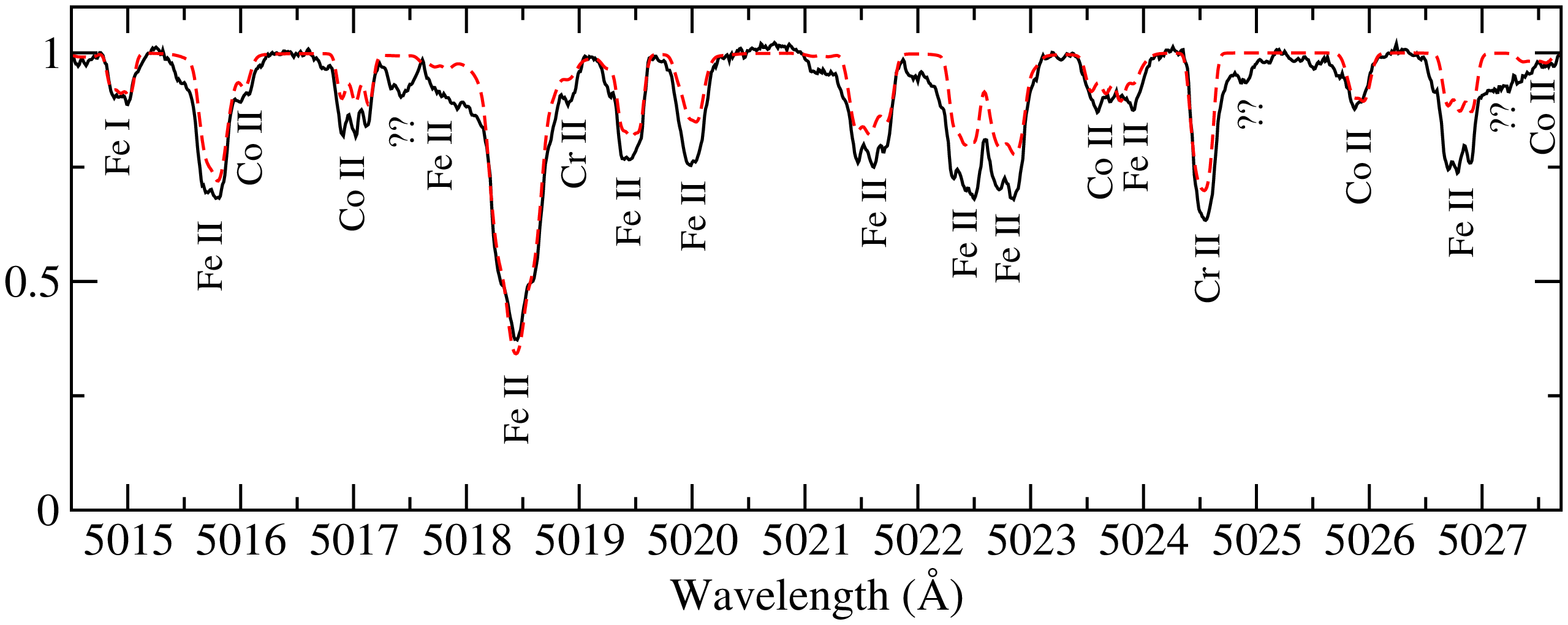}
\caption{Example of three synthesised spectral windows for the 28-04-2014 HARPSpol
spectrum of HD~94660. The observed spectrum is in black (solid line) and the model is in red (dashed line). The bottom panel illustrates the vertical stratification of Fe in the atmosphere of HD~94660, where strong lines of Fe~{\sc ii} and weak lines of Fe~{\sc i} are fit well with the abundance of Table~\ref{abund}, but the weaker lines of Fe~{\sc ii} require an enhanced abundance to be well modelled.}
\label{model-fit}
\end{figure}
\end{center}

Insignificant variability in the strengths and shapes of most spectral lines is observed among the twelve spectra available to model for HD~94660. For lines that exhibit strong Zeeman signatures (for example Nd~{\sc iii} at 6145~\AA) differences can be seen in the line structure (i.e. relative strengths of the $\sigma$ and $\pi$ components) and, in some cases, marginal changes in the line strength. However, line depth changes are limited to less than about 0.05 in intensity. Therefore, although all spectra are modelled, we report only a mean abundance for each element. Uncertainties are determined in two ways,
depending upon the number of spectral lines available to model. For
elements with multiple lines, the uncertainties are estimated from the
observed scatter between the best-fit abundances. When only one (or
few) lines are available to model, we estimate uncertainties by changing the
best-fit abundance for each element until the reduced $\chi^{2}$
deviates from best-fit models by about 1 (i.e. $\chi^{2} = \chi_{\rm
  best}^{2} + 1$). The uncertainty is then the difference between the
two abundances and corresponds approximately to a 1~$\sigma$
uncertainty. Table~\ref{abund} lists the abundances of each element, as well as the total number of lines modelled for each element. For reference, the solar abundance ratios are provided \citep{Asplund2009}. Fig.~\ref{model-fit} provides an example of the
quality of fit for three spectral windows for the HARPSpol observation taken on April 28, 2014. Below, each element is discussed individually.

\subsection{Helium}
Many lines of helium exist to derive an abundance. He~{\sc i} $\lambda$4471 and
$\lambda$5876 are predicted to be the strongest helium lines in the spectrum, but they are not detected in the available
spectra. An upper limit for helium is derived from these
regions that is at least a factor of 12 below the solar abundance. As
expected of a magnetic Ap star, HD~94660 is He-weak.

\subsection{Oxygen}
The O~{\sc i} multiplet at 6155-56-58~\AA\ is present
in all spectra and is used to derive a mean abundance. The ESPaDOnS spectrum also includes the O~{\sc i} lines at 7771-74-75~\AA,
however, this triplet suffers from non-LTE effects and is therefore
not considered. The adopted abundance is about 2.5 times less than in the Sun. 

\subsection{Magnesium}
The only line suitable for modelling is Mg~{\sc ii} at 4481~\AA. At
all phases, this line is reasonably well fit with an abundance that is
a factor of 4 below the solar ratio. 

\subsection{Silicon}
Several clean lines of Si~{\sc ii} exist throughout the spectrum
including strong lines at 5041 and 5055-56~\AA, as well as the
weaker line at 4621~\AA. When fit simultaneously, $\lambda$5041,
5055-56 require an abundance that is about 0.5~dex less than that of
$\lambda$4621. Table~\ref{abund} reports the average of these
abundances. 

We also found a line of Si~{\sc iii} at 4552~\AA. This line requires an abundance that is more
than 10 times that of the mean abundance from the Si~{\sc ii}
lines. This is exactly what was reported for a sample of mid to late B-type stars by \citet{BaileyLand2013},
who argue that the discrepancy between the abundances derived
from the first and second ionisation states of silicon is likely the
result of strong vertical stratification in the stellar
atmosphere. The different abundance values found using stronger and
weaker lines of Si~{\sc ii}, and the line of Si~{\sc iii}, indicate
that stratification is also likely in HD~94660.   

\subsection{Calcium}
Possible useful lines of calcium at 8498, 8542 and
8662~\AA\ are present in the ESPaDOnS spectrum, however, these are
blended with the Paschen lines, which cannot be calculated correctly with Zeeman at present. Therefore, only Ca~{\sc ii} at
3933~\AA\ was used to derive the final abundance of calcium. We note
that distinctly different abundances are required to adequately fit
the wings and core of this line \citep[see][where this effect is first explained for Ca]{Babel1992}. An abundance derived from the wings
of this line, which are formed deeper in the atmosphere, suggest an
abundance that is approximately solar. On the other hand, the core
(formed higher up in the atmosphere) requires an abundance that is of order 0.8~dex
below the solar abundance. This is a symptom of vertical
stratification and it would appear calcium is
strongly stratified throughout the atmosphere of HD~94660. The mean
abundance of calcium is reported in Table~\ref{abund}.

\subsection{Titanium}
For deriving the mean abundance of titanium, there are several lines
to choose throughout the blue spectrum. We have modelled the lines of
Ti~{\sc ii} at 4533, 4549, 4563, 4568 and 4571~\AA, and also at 4798 and 4805~\AA. Each
set of lines is synthesised separately and within each set the lines
are modelled simultaneously. The final abundance is the mean of these
two values (see Table~\ref{abund})

The abundance that fits well the weaker
lines of $\lambda$4568 and 4798, as well as the weak lines that are
blended with other Fe-peak elements at $\lambda$4533 and 4549, is
systematically too large for the stronger lines at $\lambda$4563,
4571 and 4805. This is a symptom of vertical stratification \citep[][]{Bagnuloetal2001}.  Nevertheless, Ti is clearly overabundant compared to the Sun by about a factor of 30. 
 
\subsection{Chromium}
Unlike the other Fe-peak elements, the abundance derived for Cr
is not as drastically dependent upon the lines modelled, which
suggests that Cr may not be significantly stratified throughout the stellar
atmosphere. This is illustrated in Fig.~\ref{model-fit} where most
lines of Cr~{\sc ii} (both strong and weak), as well as two lines of
Cr~{\sc i}, are fit well with the same uniform abundance. Note that HD~94660 is relatively hot for the presence of neutral atoms, however, these lines are unambiguously detected in all modelled spectra and are predicted to be visible by {\sc zeeman.f}. Furthermore, tests of the depth of core formation for these lines indicate that they are formed in the upper part of the stellar atmosphere, where the temperature is necessarily cooler and consistent with the presence of neutral chromium. However,
over 30 lines of Cr~{\sc ii} were tested throughout the spectrum with this uniform abundance,
and a notable fraction of weaker lines require a larger abundance
than is recorded in Table~\ref{abund} to be adequately modelled. Although apparently not as
significant as other elements, stratification may still be present for
Cr. The final abundance of Cr was found by simultaneously fitting Cr~{\sc ii} lines
at 4531, 4539, 4555, 4558, 4565, 4587,4588, 4590 and 4592~\AA. At all
phases, Cr is modelled well with an abundance that is of order 300
times larger than the solar ratio.

\subsection{Manganese}
The abundance of Mn was derived from the pair of Mn~{\sc ii} lines at
6122~\AA. The lines are well modelled at all phases with a uniform
abundance that is of order 8 times larger than the solar abundance. 

\subsection{Iron}
One mean abundance is not adequate to satisfactorily model all the
spectral lines of Fe. It is clear from Fig.~\ref{model-fit} that the
adopted abundance fits the weaker lines well, but is systematically
too large for the stronger Fe~{\sc ii} lines. This is evident
throughout the spectrum, with well over 20 Fe lines modelled with the
uniform abundance listed in Table~\ref{abund}. This is symptomatic of
vertical stratification in the atmosphere. Multiple lines of Fe~{\sc
  ii} are used to compute the mean abundance {\bf of} iron and the larger
estimated uncertainty is indicative of the discrepancies found when
modelling each line. The weaker Fe~{\sc ii} lines at 5029, 5031, 5032,
5034, 5035 and 5036 were fit simultaneously, and required an abundance larger by 0.6-0.7~dex than that deduced using strong lines such as 4583
and 5018~\AA. Three lines of Fe~{\sc i}  at 5014, 6136 and 6137~\AA\ were also
modelled and required a similar abundance to what is required to fit the
stronger Fe~{\sc ii} lines. It is clear that Fe is at
least of order 30 times more abundant than in the Sun.

\subsection{Cobalt}
The spectrum appears extraordinarily rich in Co~{\sc ii}. Several lines of Co~{\sc ii} are scattered throughout the spectrum. A total of eight lines were modelled: 5016, 5017, 5023, 5025, 5027, 5129, 5135, 5131~\AA. All lines of Co~{\sc ii} are reasonably well modelled with a uniform abundance and suggest an abundance that is at least 300 times greater than that of the Sun.

The detection of cobalt in the spectrum of a hot Ap star is rare \citep[we note that it is more common in the cooler roAp stars such as 10~Aquilae;][]{Nesvacil2013}. We are only aware of two other Co-strong hot magnetic Ap stars \citep[HR~1094 and HR~5049;][]{Nielsen2000,Dwo1980}. The abundances for both these Co-strong stars were derived without including the effects of the magnetic field and suggest abundances of this element greater than 1000 times that of the Sun. It is unclear how prevalent Co is in the atmospheres of magnetic Ap stars, which emphasises the importance of further detailed analyses of other magnetic Ap stars.

\subsection{Nickel}
Ni~{\sc ii} at 4067~\AA\ was used to derive the mean abundance. The
line is reasonably well modelled in all the spectra and Ni is apparently
overabundant compared to the solar abundance ratio by a factor of
about 3.

\subsection{Strontium}
Sr~{\sc ii} lines at 4077 and 4215~\AA\ are present in the blue spectrum at
all phases. The accurate modelling of these lines depends strongly on
the adopted abundance of Cr~{\sc ii}  with which the two Sr lines are blended.
Since the blending Cr lines are weak lines in the wings of the Sr lines, we use a somewhat
larger abundance for Cr than the value listed in Table~\ref{abund} when
modelling the lines of strontium. In this way, we get more concordant
results between these two lines of Sr~{\sc ii}, but the overall agreement is still poor. Nevertheless, it is clear that Sr is overabundant, probably by a factor of about 25 compared to the Sun.

\subsection{Lanthanum}
Two lines of La~{\sc ii} that are suitable for modelling were found in the spectrum at 4605
and 6126~\AA. The abundance was derived from the former and tested
using the latter. This rare-earth element is over 2000 times more
abundant than in the Sun. 

\subsection{Cerium}
Several lines of Ce~{\sc ii} are available to model throughout the
spectrum including 4560, 4562 and 4628~\AA. The final abundance was
found from modelling $\lambda$4628, which satisfactorily fits the
other lines of Ce. The models suggest that this rare-earth element is more
than 3000 times the solar abundance ratio. 

Interestingly, unlike what was found by \citet{Bail2013} for HD~147010, there is no
discrepancy between the abundances derived from 4560-62~\AA\
compared to 4628~\AA\ and suggests that perhaps the discrepancy they
report between these lines for that star may not be due to inaccurate $gf$
values, but instead possibly due to an unrecognised blend or
blends in this more rapidly rotating magnetic star (which has a \vsi\ of about 15~\kms). 

\subsection{Praseodymium}
Lines of Pr~{\sc iii} at 4625, 6160 and 6161~\AA\ are available to
model in all spectra. For the ESPaDOnS spectrum, we were
also able to model Pr~{\sc iii} at 7781~\AA. Little to no variation
is observed in the strength of these spectral lines and a uniform abundance models the
observed spectrum well at all phases. Similarly to the other
rare-earth elements, Pr is dramatically more abundant than in the Sun,
by a factor of order 3000.

\subsection{Neodymium}
Nd has the richest spectrum of all the rare-earth elements with
multiple lines of Nd~{\sc iii} available for modelling: 4570, 4625,
4627, 4911, 4912, 4914, 5050, 5127 and 6145~\AA. In general, abundances
derived from each line agree well with one another and are reasonably well
modelled at all phases. Nd has the highest abundance of any rare-earth element studied here, being of order 10$^{4}$ times the solar value.  

\subsection{Europium}
Two suitable lines of Eu~{\sc ii} are present for modelling: 4129 and 6645~\AA. One line of Eu~{\sc iii} is also available at 6666~\AA, however, this line is badly blended with Fe-peak elements. This rare-earth element is about 1000 times more abundant than in the Sun. 

\section{Radial Velocity Variations}\label{rv_sect}
\begin{center}
\begin{figure}
\centering
\includegraphics*[angle=0,width=0.5\textwidth]{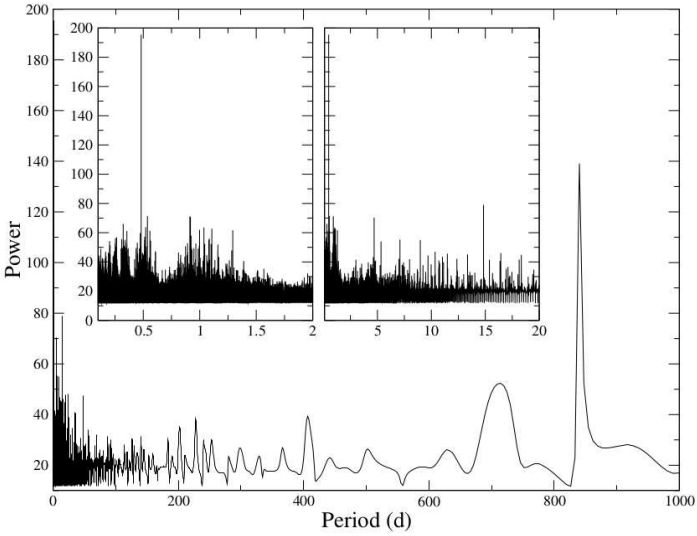}
\caption{Lomb-Scargle periodogram from the radial velocity variations of HD~94660.}
\label{periodogram}
\end{figure}
\end{center}

In the process of performing our spectroscopic analysis, it became evident
that HD~94660 exhibits significant radial velocity variations. We thus report substantial radial velocity (RV) variations in the magnetic standard star HD~94660, with a range in velocities of order 35~\kms. These variations were first reported by \citet{Mathys1997}, who noted that the orbital period should not be more than about 2 years, significantly shorter than the rotation period. Some years later, \citet{Mathys2013} determined a value of the orbital period for HD~94660 of 848.96~days. Our search for the best-fit period of these radial velocity variations was carried out using the Lomb-Scargle method \citep{press92}. The most significant frequencies in the periodogram are located at periods $\sim$0.5\,d and $\sim$840\,d, as shown in Fig.~\ref{periodogram}. The time series of observations is insufficient to provide a unique period and more observations are required in order to better constrain any periodic behaviour; however, we verify that the measurements phase well with these periods, and do not appear to vary in a coherent way when phased with other periods corresponding to lower peaks in the periodogram.   

Due to the limited temporal sampling of our dataset and the precision of our measurements, it is difficult to directly measure the expected radial velocity differences from spectra taken on the same night. Therefore, it is not possible to verify the plausibility of the $\sim$0.5\,d period solution. However, as shown by \citet{neiner12}, rapid radial velocity variations that occur over the timescale of a single polarimetric sequence can induce detectable signatures in the diagnostic null profiles. These signatures result from residual polarisation signals that were not properly cancelled during processing because of the radial velocity shifts. Therefore, the polarimetric spectra afford us the opportunity to test for the presence of short-period variations. To do so, we compared a synthetic model\footnote{This model is constructed by producing a sequence of individual synthetic sub-exposures with different radial velocities and treating them in the same fashion as the observations using the double-ratio method \citep{donati97}.}, which takes into account the predicted radial velocity variations according to the $\sim$0.5\,d period solution, to the mean LSD profiles extracted from the ESPaDOnS spectrum to test if null signatures should be present. The ESPaDOnS spectrum was obtained with the longest exposure time, and therefore should show the largest effect due to velocity shifts. Because of the strong polarisation signal, our results show that when radial velocity shifts are added to each individual sub-exposure (in agreement with the expected variation suggested by short-period variations), the resulting null profile should also show a strong signature, which is easily detectable in the mean LSD profile. While this test cannot definitively rule out the possibility of short-term radial velocity variations, it does provide compelling evidence against it. This result suggests that the $\sim$0.5\,d period is probably an alias of the correct period. We tested the plausibility of this hypothesis by subtracting a sinusoidal fit to the RV variations corresponding to the $\sim$840\,d period plus its first harmonic, and recomputed the periodogram on the residuals. The resulting periodogram no longer shows any significant power about 0.5\,d, demonstrating it to be an alias of the $\sim$840\,d period. 

If the radial velocities do vary with the $\sim$840\,d period, it is our conclusion that these variations are likely the result of binarity. If these variations were due to shifts in the centre-of-gravity of the line profile due to the inhomogeneous surface distribution (e.g. spots) that is common among Ap/Bp stars, then we would expect much smaller radial velocity shifts, which reflect the line distortion, with a maximum velocity range of the order of the line width. As well, we would expect periodicity consistent with the rotational period or one of its harmonics. 

\begin{center}
\begin{figure}
\centering
\includegraphics*[angle=0,width=0.5\textwidth]{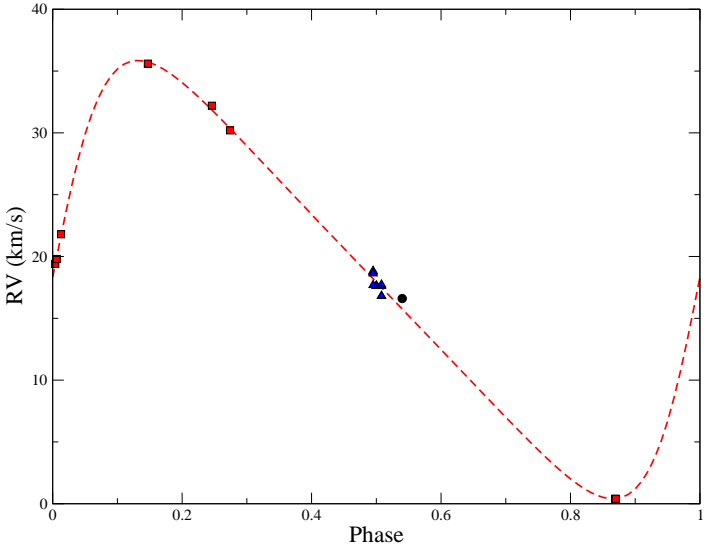}
\caption{Radial velocity orbital solution for the long 840~d period (dashed red line). Shown are the RV measurements for the HARPSpol (red squares), UVES (blue triangles) and ESPaDOnS (black dot) spectra.}
\label{rvorbit}
\end{figure}
\end{center}

If we adopt this period as the orbital period then we obtain the orbital solution given in Table~\ref{orb_params} using version 1.0.2 of the program {\sc losp} \citep[Li{\` e}ge Orbital Solution Package;][]{rauw00}. Figure~\ref{rvorbit} displays the RV orbital curve for this long period. Of particular interest, this solution suggests a high eccentricity $\sim$0.4 and a high mass-function $f(m) = m^3\sin^3 i/(M + m)^2 \sim0.4$ (where $M$ is the mass of the observed star, $m$ is the mass of the unseen companion and $i$ is the orbital inclination). Using the 1$\sigma$ limits of our estimated mass of HD~94660, this mass-function implies a lower mass limit for the companion of $\gtrsim$2~\msun. However, after co-aligning our spectra we find no evidence to suggest the presence of any additional spectral features that are not associated with HD~94660, which should be visible if HD~94660 hosted a $\sim$2~\msun\ MS companion. 
Given the relatively high mass-function permitted by our preliminary orbital solution, the most likely candidate is a high-mass neutron star or black hole; however, further data is required to constrain the period and verify the orbital solution before any definitive conclusions can be made. Furthermore, our solution does not rule out the possibility of a hierarchical system, where the companion is a combination of several objects with a total combined mass of $\gtrsim$2~\msun\ (such as two white dwarfs).

\begin{table}
\caption{Log of radial velocity variations of HD~94660. Recorded are
  the Date and JD of the observations and RVs. The
  uncertainty in the measurements of RV are less than about $\pm$1~\kms. }
\centering
\begin{tabular}{llc}
\hline\hline
 Date & JD & RV\\
(DD-MM-YYYY) & (2450000+) & (\kms)\\
\hline
01-05-2001 & 2031.464 & 16.8\\ 
01-08-2001 & 2038.441 & 16.9  \\
03-12-2005 & 3707.841 & 18.1 \\
09-01-2006 & 3745.167 & 16.6 \\
24-05-2009 & 4975.546 & 19.4 \\
25-05-2009 & 4976.536 & 19.8 \\
31-05-2009 & 4982.605 & 21.8\\
05-01-2010 & 5201.833 & 30.2\\
19-05-2011 & 5701.450 & 0.4\\
20-05-2011 & 5702.449 & 0.4\\
01-04-2012 & 6018.559 & 32.2\\
28-04-2014 & 6775.610 & 35.6\\
\hline\hline
\end{tabular}
\label{rv-var}
\end{table}

\begin{table}
\caption{Preliminary orbital parameters for HD~94660. Included are the orbital period ($P$), the time of periastron ($T_0$; 2450000+), the centre-of-mass velocity ($\gamma$), the eccentricity ($e$), the velocity semi-amplitude ($K$), the angle of the line of nodes ($\omega$), the projected semi-major axis ($a\sin i$), the mass-function ($f(m)$) and the standard deviation fit ($\sigma$).}
\centering
\begin{tabular}{cc}
\hline\hline
 Parameter & Best-fit\\
\hline
$P$ (d) & 840* \\
$T_0$ & $1611\pm3$\\
$\gamma$ (km\,s$^{-1}$) & $18\pm0.1$\\
$e$ & $0.38\pm0.03$\\
$K$ (km\,s$^{-1}$) & $17.7\pm0.3$\\
$\omega$ (deg) & $271\pm2$ \\
$a\sin i$ (R$_\odot$) & $272\pm6$\\
$f(m)$ ($M_{\odot}$) & $0.39\pm0.02$ \\
$\sigma$ & 0.41\\
\hline\hline
\multicolumn{2}{c}{* - fixed}\\
\end{tabular}
\label{orb_params}
\end{table}

\section{Discussion and Conclusions}
HD~94660 is an Ap star commonly used as a magnetic standard for polarimetric observations in the southern hemisphere. It has an effective temperature \te\ = 11\,300~K with $ \log L/L_{\odot} =$ 2.02 and mass $M/M_{\odot} =$ 3.0. The rotation period is approximately 2800~d and \vsi\ is less than about 2~\kms. The surface magnetic field strength is of order 6~kG globally.

The aim of this project is to establish a preliminary magnetic field model of HD~94660 to use to estimate the atmospheric abundances of several chemical elements. The magnetic field model adopted is a simple, low-order axisymmetric multipole expansion whose parameters are established by fitting the observed periodic variations in \bz\ and \bs\ to computed models. This model is produced in the framework of the oblique rotator model and reasonably reproduces the observed variations in \bz\ and \bs\ with rotational phase (Fig.~\ref{magfield}). The model is only a coarse approximation to the true field geometry of HD~94660, but is able to reproduce the observed Zeeman splitting and Stokes $V$ signatures with rotational phase reasonably well (Fig.~\ref{model}). This simple magnetic model is adequate to make it possible to determine a first approximation of the atmospheric abundances of this star. The actual parameters of this model are discussed in Sect.~4.3 and presented in Table~1.

We have used a dozen high-dispersion $I$ spectra, well distributed in phase over the rotation cycle of the star, for a preliminary investigation of the surface chemistry and a characterisation of how the derived abundances may vary over the stellar surface. From the magnetic field model, the fact that $i$ plus $\beta$ is small ($\ltsimeq 50^{\degr}$) indicates that we are mainly observing one magnetic hemisphere, with limited information about the opposite magnetic hemisphere. Although more than half of the stellar surface is seen, our investigations indicate very little abundance variations with rotational phase. For all elements studied, a single abundance fits well all available spectra and therefore a model with a uniform abundance distribution over the stellar surface is adopted.

As is expected for magnetic Ap stars, most elements studied have non-solar abundances. The abundances of O, Mg and Ca are all slightly below solar abundance ratios. Only an upper limit for He is possible which clearly classifies HD~94660 as He-weak, with a value at least 10 times less than the solar abundance. All other elements studied are more abundant than in the Sun. Most drastically, the rare-earth elements La, Ce, Pr and Nd are all between about 10$^{3}$ to 10$^{4}$ times more abundant than in the Sun. The abundances of Si, Sr and the Fe-peak elements Ti, Cr, Mn, Fe, Co and Ni are also larger than the solar abundance ratios (by factors of order 10$^{3}$ or less).  Although no significant variations with co-latitude are observed in the stellar spectra, within a single spectrum the Fe-peak elements, calcium and silicon exhibit strong evidence of vertical stratification. This is most notable for Fe, where abundances derived from weak and strong lines of Fe~{\sc ii} (as well as weaker neutral Fe lines) differ by factors of order 10. Furthermore, the discrepant abundances derived from lines of Si~{\sc ii} and Si~{\sc iii} reported by \citet{BaileyLand2013} are also present in HD~94660. A more detailed analysis of abundance stratification in this star is clearly warranted. The discovery of Co in HD~94660 was surprising and further detailed studies of other hot magnetic Ap stars are recommended to ascertain the overabundance of this element in the atmospheres of these stars.

\citet{LBF2014} highlight the fact that HD~94660 is a star where there is poor agreement between the \bz\ measurements made from instruments with lower and higher resolutions. This is evident in Fig.~\ref{magfield} where the field strengths extracted from the HARPSpol and ESPaDOnS spectra using the entire metallic spectrum disagree with the FORS1 measurements. This phenomenon is also present in other stars such as HD~318107 \citep[see][]{Baileyetal2011}, NGC~2169-12, NGC~2244-334 and HD~149277 \citep[see][]{paper1}. The disagreement between field measurements, using LSD, of different elements found in HD~94660 is also not uncommon (e.g. \citet[][]{MM2000,Baileyetal2011} for HD~318107 and \citet[][]{Baileyetal2012} for HD~133880). These types of discrepancies are generally considered an indication of the inhomogeneous field distributions and large horizontal abundance variations (``spots'') on the stellar surface. This hypothesis is supported by more detailed maps of Ap stars using magnetic Doppler imaging (MDI) in which clear, complex field distributions and anomalous abundance spots are observed on the stellar surface \citep[e.g.][]{Kochetal2004}.

The RV variations measured in HD~94660 arise from orbital motion. Because of the long period currently favoured by the measurements ($\sim$840~d), this would suggest a massive compact companion such as a high-mass neutron star or black hole, or possibly a hierarchical system, where the companion is a combination of several objects with a total combined mass of $\gtrsim$2~\msun. It is rare for an A--type star to host a massive compact companion \citep[see][]{Kaper2006} and further monitoring is warranted to better constrain the properties of the companion. This system could help to establish the role that binarity may play in the origin of magnetism in stars with radiative envelopes \citep[e.g.][ and references therein]{grunhut14}. 

HD~94660 is a star that warrants further investigation and highlights the need to study more sharp-lined magnetic stars in detail. Such studies are crucial to further understand the interplay between the magnetic field and the formation of vertical stratification in the atmospheres of Ap stars. They also provide important laboratories to test our multi-line techniques for measuring magnetic fields, such as LSD, by allowing measurement of magnetic field strengths from individual lines, a task that is not possible for stars that are fast rotators. At present, it is unclear what the discrepant field measurements for different lines of the same element are telling us about the chemical or magnetic structure of Ap stars, and therefore further analysis is required. Long term monitoring is also clearly indicated to firmly establish the nature of the RV variations observed in HD~94660.

\begin{acknowledgements}
{The authors thank Dr. Stephan Geier of ESO for helpful discussions. The authors also thank R.H.D. Townsend for the Lomb-Scargle code. JDL acknowledges financial support from the Natural Sciences and Engineering Research Council of Canada. The authors also thank the referee Gautier Mathys for his comments that helped improve the manuscript. }
\end{acknowledgements}

\bibliographystyle{aa}
\bibliography{BGL2014}

\begin{thebibliography}{52}
\expandafter\ifx\csname natexlab\endcsname\relax\def\natexlab#1{#1}\fi

\bibitem[{{Asplund} {et~al.}(2009){Asplund}, {Grevesse}, {Sauval}, \&
  {Scott}}]{Asplund2009}
{Asplund}, M., {Grevesse}, N., {Sauval}, A.~J., \& {Scott}, P. 2009, ARA\&A,
  47, 481

\bibitem[{{Babel}(1992)}]{Babel1992}
{Babel}, J. 1992, A\&A, 258, 449

\bibitem[{{Bagnulo} {et~al.}(2006){Bagnulo}, {Landstreet}, {Mason}, {Andretta},
  {Silaj}, \& {Wade}}]{paper1}
{Bagnulo}, S., {Landstreet}, J.~D., {Mason}, E., {et~al.} 2006, A\&A, 450, 777

\bibitem[{{Bagnulo} {et~al.}(2001){Bagnulo}, {Wade}, {Donati}, {Landstreet},
  {Leone}, {Monin}, \& {Stift}}]{Bagnuloetal2001}
{Bagnulo}, S., {Wade}, G.~A., {Donati}, J.-F., {et~al.} 2001, A\&A, 369, 889

\bibitem[{{Bailey}(2014)}]{Bailey2014}
{Bailey}, J.~D. 2014, A\&A, 568, A38

\bibitem[{{Bailey} {et~al.}(2012){Bailey}, {Grunhut}, {Shultz}, {Wade},
  {Landstreet}, {Bohlender}, {Lim}, {Wong}, {Drake}, \&
  {Linsky}}]{Baileyetal2012}
{Bailey}, J.~D., {Grunhut}, J., {Shultz}, M., {et~al.} 2012, MNRAS, 423, 328

\bibitem[{{Bailey} \& {Landstreet}(2013{\natexlab{a}})}]{BaileyLand2013}
{Bailey}, J.~D. \& {Landstreet}, J.~D. 2013{\natexlab{a}}, A\&A, 551, A30

\bibitem[{{Bailey} \& {Landstreet}(2013{\natexlab{b}})}]{Bail2013}
{Bailey}, J.~D. \& {Landstreet}, J.~D. 2013{\natexlab{b}}, MNRAS, 432, 1687

\bibitem[{{Bailey} {et~al.}(2014){Bailey}, {Landstreet}, \&
  {Bagnulo}}]{BLB2014}
{Bailey}, J.~D., {Landstreet}, J.~D., \& {Bagnulo}, S. 2014, A\&A, 561, A147

\bibitem[{{Bailey} {et~al.}(2011){Bailey}, {Landstreet}, {Bagnulo}, {Fossati},
  {Kochukhov}, {Paladini}, {Silvester}, \& {Wade}}]{Baileyetal2011}
{Bailey}, J.~D., {Landstreet}, J.~D., {Bagnulo}, S., {et~al.} 2011, A\&A, 535,
  A25

\bibitem[{{Bohlender} {et~al.}(1993){Bohlender}, {Landstreet}, \&
  {Thompson}}]{BLT1993}
{Bohlender}, D.~A., {Landstreet}, J.~D., \& {Thompson}, I.~B. 1993, A\&A, 269,
  355

\bibitem[{{Borra} \& {Landstreet}(1975)}]{BL1975}
{Borra}, E.~F. \& {Landstreet}, J.~D. 1975, PASP, 87, 961

\bibitem[{{Donati} \& {Landstreet}(2009)}]{DL2009}
{Donati}, J.-F. \& {Landstreet}, J.~D. 2009, ARA\&A, 47, 333

\bibitem[{{Donati} {et~al.}(1997){Donati}, {Semel}, {Carter}, {Rees}, \&
  {Collier Cameron}}]{donati97}
{Donati}, J.-F., {Semel}, M., {Carter}, B., {Rees}, D., \& {Collier Cameron},
  A. 1997, MNRAS, 291, 658

\bibitem[{{Dworetsky} {et~al.}(1980){Dworetsky}, {Trueman}, \&
  {Stickland}}]{Dwo1980}
{Dworetsky}, M.~M., {Trueman}, M.~R.~G., \& {Stickland}, D.~J. 1980, A\&A, 85,
  138

\bibitem[{{Girardi} {et~al.}(2000){Girardi}, {Bressan}, {Bertelli}, \&
  {Chiosi}}]{Girardi2000}
{Girardi}, L., {Bressan}, A., {Bertelli}, G., \& {Chiosi}, C. 2000, A\&AS, 141,
  371

\bibitem[{{Grunhut} \& {Alecian}(2014)}]{grunhut14}
{Grunhut}, J.~H. \& {Alecian}, E. 2014, in IAU Symposium, Vol. 302, IAU
  Symposium, 70--79

\bibitem[{{Hensberge}(1993)}]{Hensberge1993}
{Hensberge}, H. 1993, in Astronomical Society of the Pacific Conference Series,
  Vol.~44, IAU Colloq. 138: Peculiar versus Normal Phenomena in A-type and
  Related Stars, ed. M.~M. {Dworetsky}, F.~{Castelli}, \& R.~{Faraggiana}, 547

\bibitem[{{Kaper} {et~al.}(2006){Kaper}, {van der Meer}, {van Kerkwijk}, \&
  {van den Heuvel}}]{Kaper2006}
{Kaper}, L., {van der Meer}, A., {van Kerkwijk}, M., \& {van den Heuvel}, E.
  2006, The Messenger, 126, 27

\bibitem[{{Kochukhov} {et~al.}(2004){Kochukhov}, {Bagnulo}, {Wade}, {Sangalli},
  {Piskunov}, {Landstreet}, {Petit}, \& {Sigut}}]{Kochetal2004}
{Kochukhov}, O., {Bagnulo}, S., {Wade}, G.~A., {et~al.} 2004, A\&A, 414, 613

\bibitem[{{Kochukhov} {et~al.}(2010){Kochukhov}, {Makaganiuk}, \&
  {Piskunov}}]{kochukhov10}
{Kochukhov}, O., {Makaganiuk}, V., \& {Piskunov}, N. 2010, A\&A, 524, A5

\bibitem[{{Kunzli} {et~al.}(1997){Kunzli}, {North}, {Kurucz}, \&
  {Nicolet}}]{geneva}
{Kunzli}, M., {North}, P., {Kurucz}, R.~L., \& {Nicolet}, B. 1997, A\&AS, 122,
  51

\bibitem[{{Kupka} {et~al.}(1999){Kupka}, {Piskunov}, {Ryabchikova}, {Stempels},
  \& {Weiss}}]{vald4}
{Kupka}, F., {Piskunov}, N.~E., {Ryabchikova}, T.~A., {Stempels}, N.~C., \&
  {Weiss}, W.~W. 1999, \aap, 138, 119

\bibitem[{{Kupka} {et~al.}(2000){Kupka}, {Ryabchikova}, {Piskunov}, {Stempels},
  \& {Weiss}}]{vald1}
{Kupka}, F.~G., {Ryabchikova}, T.~A., {Piskunov}, N.~E., {Stempels}, H.~C., \&
  {Weiss}, W.~W. 2000, Baltic Astronomy, 9, 590

\bibitem[{{Landstreet}(1988)}]{Landstreet1988}
{Landstreet}, J.~D. 1988, ApJ, 326, 967

\bibitem[{{Landstreet} {et~al.}(2007){Landstreet}, {Bagnulo}, {Andretta},
  {Fossati}, {Mason}, {Silaj}, \& {Wade}}]{paper2}
{Landstreet}, J.~D., {Bagnulo}, S., {Andretta}, V., {et~al.} 2007, A\&A, 470,
  685

\bibitem[{{Landstreet} {et~al.}(2014){Landstreet}, {Bagnulo}, \&
  {Fossati}}]{LBF2014}
{Landstreet}, J.~D., {Bagnulo}, S., \& {Fossati}, L. 2014, A\&A, 572, A113

\bibitem[{{Landstreet} \& {Mathys}(2000)}]{LM2000}
{Landstreet}, J.~D. \& {Mathys}, G. 2000, A\&A, 359, 213

\bibitem[{{Landstreet} {et~al.}(2008){Landstreet}, {Silaj}, {Andretta},
  {Bagnulo}, {Berdyugina}, {Donati}, {Fossati}, {Petit}, {Silvester}, \&
  {Wade}}]{paper3}
{Landstreet}, J.~D., {Silaj}, J., {Andretta}, V., {et~al.} 2008, A\&A, 481, 465

\bibitem[{{Makaganiuk} {et~al.}(2011){Makaganiuk}, {Kochukhov}, {Piskunov},
  {Jeffers}, {Johns-Krull}, {Keller}, {Rodenhuis}, {Snik}, {Stempels}, \&
  {Valenti}}]{makaganiuk11}
{Makaganiuk}, V., {Kochukhov}, O., {Piskunov}, N., {et~al.} 2011, \aap, 525,
  A97

\bibitem[{{Manfroid} \& {Mathys}(2000)}]{MM2000}
{Manfroid}, J. \& {Mathys}, G. 2000, A\&A, 364, 689

\bibitem[{{Mathys}(1990)}]{Mathys1990}
{Mathys}, G. 1990, A\&A, 232, 151

\bibitem[{{Mathys}(2013)}]{Mathys2013}
{Mathys}, G. 2013, in Astronomical Society of the Pacific Conference Series,
  Vol. 479, Progress in Physics of the Sun and Stars: A New Era in Helio- and
  Asteroseismology, ed. H.~{Shibahashi} \& A.~E. {Lynas-Gray}, 81

\bibitem[{{Mathys} \& {Hubrig}(1997)}]{Mathys+Hubrig1997}
{Mathys}, G. \& {Hubrig}, S. 1997, A\&AS, 124, 475

\bibitem[{{Mathys} {et~al.}(1997){Mathys}, {Hubrig}, {Landstreet}, {Lanz}, \&
  {Manfroid}}]{Mathys1997}
{Mathys}, G., {Hubrig}, S., {Landstreet}, J.~D., {Lanz}, T., \& {Manfroid}, J.
  1997, A\&AS, 123, 353

\bibitem[{{Napiwotzki} {et~al.}(1993){Napiwotzki}, {Schoenberner}, \&
  {Wenske}}]{NSW}
{Napiwotzki}, R., {Schoenberner}, D., \& {Wenske}, V. 1993, A\&A, 268, 653

\bibitem[{{Neiner} {et~al.}(2012){Neiner}, {Landstreet}, {Alecian}, {Owocki},
  {Kochukhov}, {Bohlender}, \& {MiMeS Collaboration}}]{neiner12}
{Neiner}, C., {Landstreet}, J.~D., {Alecian}, E., {et~al.} 2012, \aap, 546, A44

\bibitem[{{Nesvacil} {et~al.}(2013){Nesvacil}, {Shulyak}, {Ryabchikova},
  {Kochukhov}, {Akberov}, \& {Weiss}}]{Nesvacil2013}
{Nesvacil}, N., {Shulyak}, D., {Ryabchikova}, T.~A., {et~al.} 2013, A\&A, 552,
  A28

\bibitem[{{Nielsen} \& {Wahlgren}(2000)}]{Nielsen2000}
{Nielsen}, K. \& {Wahlgren}, G.~M. 2000, A\&A, 356, 146

\bibitem[{{Piskunov} {et~al.}(2011){Piskunov}, {Snik}, {Dolgopolov},
  {Kochukhov}, {Rodenhuis}, {Valenti}, {Jeffers}, {Makaganiuk}, {Johns-Krull},
  {Stempels}, \& {Keller}}]{piskunov11}
{Piskunov}, N., {Snik}, F., {Dolgopolov}, A., {et~al.} 2011, The Messenger,
  143, 7

\bibitem[{{Piskunov} {et~al.}(1995){Piskunov}, {Kupka}, {Ryabchikova}, {Weiss},
  \& {Jeffery}}]{vald3}
{Piskunov}, N.~E., {Kupka}, F., {Ryabchikova}, T.~A., {Weiss}, W.~W., \&
  {Jeffery}, C.~S. 1995, \aap, 112, 525

\bibitem[{{Piskunov} \& {Valenti}(2002)}]{piskunov02}
{Piskunov}, N.~E. \& {Valenti}, J.~A. 2002, A\&A, 385, 1095

\bibitem[{{Press} {et~al.}(1992){Press}, {Teukolsky}, {Vetterling}, \&
  {Flannery}}]{press92}
{Press}, W., {Teukolsky}, S., {Vetterling}, W., \& {Flannery}, B. 1992,
  {Numerical recipes in FORTRAN. The art of scientific computing} (Cambridge:
  Cambridge University Press)

\bibitem[{{Rauw} {et~al.}(2000){Rauw}, {Sana}, {Gosset}, {Vreux}, {Jehin}, \&
  {Parmentier}}]{rauw00}
{Rauw}, G., {Sana}, H., {Gosset}, E., {et~al.} 2000, \aap, 360, 1003

\bibitem[{{Rees} \& {Semel}(1979)}]{rees79}
{Rees}, D. \& {Semel}, M. 1979, A\&A, 74, 1

\bibitem[{{Renson} {et~al.}(1991){Renson}, {Gerbaldi}, \&
  {Catalano}}]{Rensonetal1991}
{Renson}, P., {Gerbaldi}, M., \& {Catalano}, F.~A. 1991, A\&AS, 89, 429

\bibitem[{{Rusomarov} {et~al.}(2013){Rusomarov}, {Kochukhov}, {Piskunov},
  {Jeffers}, {Johns-Krull}, {Keller}, {Makaganiuk}, {Rodenhuis}, {Snik},
  {Stempels}, \& {Valenti}}]{rusomarov13}
{Rusomarov}, N., {Kochukhov}, O., {Piskunov}, N., {et~al.} 2013, A\&A, 558, A8

\bibitem[{{Ryabchikova}(1991)}]{Ryab1991}
{Ryabchikova}, T.~A. 1991, in IAU Symposium, Vol. 145, Evolution of Stars: the
  Photospheric Abundance Connection, ed. G.~{Michaud} \& A.~V. {Tutukov}, 149

\bibitem[{{Ryabchikova} {et~al.}(1997){Ryabchikova}, {Piskunov}, {Kupka}, \&
  {Weiss}}]{vald2}
{Ryabchikova}, T.~A., {Piskunov}, N.~E., {Kupka}, F., \& {Weiss}, W.~W. 1997,
  Baltic Astronomy, 6, 244

\bibitem[{{Stibbs}(1950)}]{Stibbs1950}
{Stibbs}, D.~W.~N. 1950, MNRAS, 110, 395

\bibitem[{{van Leeuwen}(2007)}]{vanLee2007}
{van Leeuwen}, F. 2007, A\&A, 474, 653

\bibitem[{{Wade} {et~al.}(2000){Wade}, {Donati}, {Landstreet}, \&
  {Shorlin}}]{wade00}
{Wade}, G.~A., {Donati}, J.-F., {Landstreet}, J.~D., \& {Shorlin}, S.~L.~S.
  2000, MNRAS, 313, 823

\end{thebibliography}
\end{document}